\documentclass[aps,pra,superscriptaddress,twocolumn]{revtex4-1}

\usepackage{amsmath,amsfonts,color,graphicx,fullpage, subfig,setspace, floatrow,dcolumn}
\newfloatcommand{capbtabbox}{table}[][\FBwidth]
\usepackage[colorlinks={true},pdftex]{hyperref}
\usepackage[singlelinecheck=off,justification=raggedright]{caption}
\hypersetup{citecolor={blue}, filecolor={blue}, linkcolor={blue}, urlcolor={blue}}

\begin{document}
\date{\today}

\author{David Hocker}
\affiliation{Department of Chemistry, Princeton University, Princeton, NJ 08544, USA}
\author{Julia Yan}
\affiliation{Operations Research Center, Massachusetts Institute of Technology, Cambridge, MA 02139}
\affiliation{Department of Chemistry, Princeton University, Princeton, NJ 08544, USA}
\author{Herschel Rabitz}
\affiliation{Department of Chemistry, Princeton University, Princeton, NJ 08544, USA}

\title{Optimal nonlinear coherent mode transitions in Bose-Einstein Condensates utilizing spatio-temporal controls}
\begin{abstract}
Bose-Einstein condensates (BECs) offer the potential to examine quantum behavior at large length and time scales, as well as forming promising candidates for quantum technology applications. Thus, the manipulation of BECs using control fields is a topic of prime interest. We consider BECs in the mean field model of the Gross-Pitaevskii equation (GPE), which contains linear and nonlinear features, both of which are subject to control. In this work we report successful optimal control simulations of a one dimensional GPE by modulating the  linear and nonlinear terms to stimulate  transitions into excited coherent modes. The linear and nonlinear controls are allowed to freely vary over space and time to seek their optimal forms. The determination of the excited coherent modes targeted for optimization is numerically performed through an adaptive imaginary time propagation method. Numerical simulations are performed for optimal control of mode-to-mode transitions between the ground coherent mode and excited modes of a BEC trapped in a harmonic well. The results show greater than 99$\%$ success for nearly all trials utilizing reasonable initial guesses for the controls, and analysis of the optimal controls reveals primarily direct transitions between initial and target modes. The success of using solely the nonlinearity term as a control opens up further research toward exploring novel control mechanisms inaccessible to linear Schr\"odinger-type systems.
\end{abstract}

\maketitle

\section{Introduction}
The control of quantum phenomena spans a wide range of time and length scales \cite{brif_jnewphys_2010, quantphyslett_1_1_2012}. The large dynamical scale of Bose-Einstein condensates (BECs) makes them ideal systems for studying macroscopic quantum effects \cite{Pethick_monograph, Kevrekidis_monograph}, in addition to potential quantum technology applications such as quantum information processing \cite{bec_qip}. Moreover, BECs also offer the possibility to explore the control of nonlinear Schr\"odinger equations (NLSE). The Gross-Pitaevskii equation (GPE) is one such NLSE, and it is utilized extensively to study dynamical features of a BEC \cite{Pethick_monograph},
\begin{align}
\label{eq:GPE_1D}
i\hbar & \frac{\partial \psi(x,t)}{\partial t}  = H(x,t) \psi(x,t) \nonumber \\
 	&=\left [H_0 + V(x,t) + g(x,t) |\psi(x,t)|^2 \right] \psi(x,t).
\end{align}
The wavefunction (order parameter) in the one-dimensional GPE shown in eq. (\ref{eq:GPE_1D}) is normalized to the number of particles in the condensate. The GPE provides a mean-field description of the condensate where the effective Hamiltonian contains the kinetic energy term $H_0$ and trapping potential $V$, accompanied by a nonlinear term $g|\psi|^2$ accounting for the weak interparticle interactions. Additionally \emph{portions} of both $V(x,t)$ and $g(x,t)$ can serve as controls (\emph{i.e.}, later referred to as $V_{\mathrm{cont}}(x,t)$ and $g_{\mathrm{cont}}(x,t)$).

The character of each quantum control problem is revealed by its landscape, defined as the physical objective as a functional of the controls. The success of a growing number of optimal control experiments, as well as vast numbers of simulations, led to the formulation of a key theorem referred to here as the ``landscape principle." The principle states that upon satisfaction of assumptions (sufficient conditions) about the controllability, surjectivity and available resources, the topology of quantum control landscapes for systems with a finite number of states allows for facile determination of optimal controls \cite{Hsieh_pra_2008, Hsieh_jcp_2009, PhysRevA.74.012721}.

The landscape principle has only been rigorously established and examined for quantum systems driven by a Schr\"odinger equation of finite dimensions that is linear with respect to the wavefunction $\psi$. While investigation into the landscape for a discrete, finite-level representation of NLSE systems suggests that the landscape principle still holds \cite{yan_pra_2013}, a numerical investigation of control of the NLSE in a spatial representation of the quantum system has yet to be performed. Although a full investigation of the landscape principle has not been performed for the GPE, there is encouraging evidence of its successful optimal control. In particular, optimal control simulations have been performed with magnetic fields described by small sets of control parameters for effective splitting of condensates and number squeezing \cite{Hohenester_pra_2007, Hohenester_pra_2009, Hohenester_compphyscomm_2014}. Additionally, an experimental implementation of these parametrized control methods has been used to create stable matter-wave optics for low-temperature interferometry experiments \cite{becwings}. 

Here we depart from the use of parametrized controls in order to examine the full, unconstrained capability of controls identified through optimization. Thus, control \emph{functions} are considered, which permit adjusting of their spatio-temporal form throughout an unconstrained optimization, thereby opening up study of the most flexible possible controls. In this context, we also consider the nonlinearity strength $g(x,t)$ in eq. (\ref{eq:GPE_1D}) as a control through its magnetic field dependence via Feshbach resonance \cite{Chin_revmodphys_2010}. In addition to this study providing a simulation test of the landscape principle for the continuous GPE, the utilization of a functional nonlinear control also opens up control strategies inaccessible to linear Schr\"odinger quantum systems.

The objective studied here is the transition between stationary modes of the GPE,
\begin{align}
\label{eq:gpe_pif}
P_{0\rightarrow f}  &=  \left |\int_{-\infty}^{\infty} \phi^*_f(x) \psi(x,T) dx \right |^2, \quad \psi(x,0) = \phi_0(x).
\end{align}
The terminology of a mode versus that of a state is associated with the stationary GPE posing a nonlinear eigenvalue or eigenfunction (mode) problem, which will be discussed further in Section \ref{sec:bectopomodes}. In particular, the present work considers the goal of maximizing the transition from the ground mode $\phi_0$  to a target excited mode $\phi_f$,
\begin{align}
\label{eq:gpe_maxpif}
\max_{V_{\text{cont}},g_{\text{cont}}} P_{0\rightarrow f}
\end{align}
at final time $T$ through the utilization of the controls $V_{\text{cont}}(x,t)$ and/or $g_{\text{cont}}(x,t)$. Maximizing state transitions is a well-studied objective in quantum optimal control of the linear Schr\"odinger equation, where typically the initial and final states are taken to be eigenstates of the field-free Hamiltonian \cite{Rabitz_pra_2006}. As mentioned above for the GPE, though, locating such eigenmodes of the control-free Hamiltonian (\emph{i.e.}, with $V$ being just a spatially dependent trapping potential and $g$ being a constant) involves solving a nonlinear eigenvalue problem, which can lead to a diversity of solutions \cite{Lieb_monograph}. One such class of eigenmodes considered here manifest coherence across the entire BEC (See the discussion in Sec. \ref{sec:bectopomodes}). Analogous to the eigenstates of the linear Schr\"odinger equation for a harmonic oscillator or particle in a box, these coherent stationary modes $\phi_j$ of energy $E_j$ correspond to a particular coherence over the entire condensate. Thus, full maximization of eq. (\ref{eq:gpe_maxpif}) would permit patterning of a BEC according to the spatial shape of $\phi_f(x)$, or possibly the shape created by a superposition of modes at time $T$ \cite{PhysRevA.56.4845, Ramos_pra_2008, PhysRevA.66.043602}. As analytical descriptions of $\phi_0$ and $\phi_f$ are generally not available, special numerical methods must be used to identify the coherent modes.

This work is organized as follows. Section \ref{sec:gp} formulates the GPE functional control problem, and section \ref{sec:becprop} details the numerical methods involved in locating control solutions. An algorithm for determining coherent modes of the GPE is presented in section \ref{sec:bectopomodes}, which is then utilized for a BEC in a harmonic trapping potential for various nonlinearity strengths. Optimal control of mode-to-mode transitions is presented in section \ref{sec:becoct}. Analysis of the identified controls is performed in section \ref{sec:bec_analysis}, followed by concluding remarks in section \ref{sec:becconclusion}. 

\section{Control of the Gross-Pitaevskii equation}
\label{sec:gp}
The optimal control of the GPE will be considered in the one-dimensional spatial representation of the dynamics given by eq. (\ref{eq:GPE_1D}). In order to highlight the controls of interest, eq. (\ref{eq:GPE_1D}) is written more explicitly for a BEC trapped in a harmonic potential well centered at position $L/2$
\begin{align}
\label{eq:GPE_1D_expl}
i\hbar \frac{\partial \psi(x,t)}{\partial t}  &= \left [H_0 + V(x,t) + g(x,t) |\psi(x,t)|^2 \right] \psi(x,t) \nonumber \\
 	&= [H_0 + \frac{m\omega^2}{2} \left (x-\frac{L}{2} \right)^2 + g_0 |\psi(x,t)|^2+ \nonumber\\
 	&  V_{\text{cont}}(x,t) + g_{\text{cont}}(x,t) |\psi(x,t)|^2   \psi(x,t),
 	\end{align}
 	where
 	\begin{align}
 	\label{eq:GPE_1D_potterms}
 	V(x,t) &= \frac{m\omega^2}{2} \left (x-\frac{L}{2} \right)^2 + V_{\text{cont}}(x,t) \\
 	g(x,t) &= (g_0 +g_{\text{cont}}(x,t)).
\end{align}
Here a harmonic well of frequency $\omega$ contains the condensate, where $m$ is the mass of individual bosons in the condensate, and $g_0$ is a characteristic, constant nonlinearity strength. For simplicity we will later set $\omega=m=1$, and normalize $\psi$ to 1. In addition to the potential well and $g_0$, eq. (\ref{eq:GPE_1D_expl}) also contains two functional control terms: a potential control $V_{\text{cont}}(x,t)$ and a ``nonlinearity control" $g_{\text{cont}}(x,t)$,  where the latter attempts to manipulate the condensate through the contribution from $|\psi(x,t)|^2$.

 The nonlinearity terms, $[g_0 +g_{\text{cont}}(x,t)]|\psi(x,t)|^2$ relate to the the scattering length of interparticle interactions inside the condensate \cite{Pethick_monograph}. In practice the scattering length has a dependence $\sim 1/(B-B_0)$  arising from an applied magnetic field $B$ (or the magnetic field component of an optical field) through a Feshbach resonance near point $B_0$ specific to the bosons forming the condensate and system setup. Here $g_0$ accounts for the intrinsic nonlinearity of the condensate, analogous to a condensate in the absence of additional magnetic fields apart from the trapping potential, and hence away from a Feshbach resonance point. $g_{\text{cont}}(x,t)$ conversely accounts for changes in the scattering length through its influence on a neighboring Feshbach resonance accessed through an applied magnetic field $B$. $g_{\text{cont}}(x,t)$ is considered tuneable along with the linear control term $V_{\text{cont}}(x,t)$.  Recent work has demonstrated spatial shaping of the nonlinearity through an optically induced Feshbach resonance control that locally modulates the nonlinearity while leaving the trapping potential unaltered \cite{PhysRevLett.115.155301}. We will more generally explore controls capable of both spatial and temporal tuning to perform optimal mode transitions in BECs.

The necessary equations for optimizing  eq. (\ref{eq:gpe_pif}) can be derived with the Lagrange multiplier method, whereby a constraint is added to eq. (\ref{eq:gpe_pif}) such that the GPE is satisfied. Differentiation of this augmented cost leads to optimality conditions and the gradient of $P_{0 \rightarrow f}$ \cite{Hohenester_pra_2007}. The optimality equations are briefly summarized below, and their derivation is given in the Supplementary Material. Introducing the constraint through a Lagrange multiplier state $p^*(x,t)$ in eq. (\ref{eq:gpe_pif}), the modified cost functional $J$ is expressed as  \cite{Hohenester_compphyscomm_2014}
\begin{align}
\label{eq:L}
& J[\psi,p,V, g]  \nonumber =  \left |\int \phi_f^{*}(x)\psi(x,T) dx \right |^2 + \nonumber \\
& 2 \text{Re} \left ( \int_0^T dt \int_{-\infty}^{\infty} dx  p^{*}(x,t)  \left [-i\hbar \frac{\partial}{\partial t}  + H(x,t) \right ]\psi(x,t) \right ).
\end{align}
The $(x,t)$-dependence may later be left as implicitly understood where no confusion will arise. The variational cost functional $J$ leads to optimality conditions that must be satisfied. These conditions are found by setting the functional derivatives of $J$  with respect to $p$ and $\psi$ equal to zero (and their complex conjugates, not shown. See Supplementary Material for full details) to respectively produce
\begin{align}
\label{eq:dL_dlambda_result_final}
i\hbar \frac{\partial}{\partial t}\psi(x,t) &=  H(t) \psi(x,t), \quad \psi(x,0) = \phi_0(x) \\
\label{eq:delL_delpsi_results}
 i\hbar  \frac{\partial}{\partial t}p(x,t) &=  \left [H(t)  + g(x,t)  |\psi(x,t)|^2 \right ] p(x,t) \nonumber \\
 &+ g(x,t) \psi^2(x,t) p^*(x,t) ,  \\
\label{eq:delL_delpsi_init} 
 p(x,T) &= -\frac{i}{\hbar}\phi_f(x)  \int_{-\infty}^{\infty} \phi_f^*(x) \psi(x,T) dx .
 \end{align}
Additionally, the derivatives with respect to the controls are
\begin{align}
\label{eq:delL_V_result_final}
\frac{\delta J}{\delta V_{\text{cont}}(x,t)} & \equiv \frac{\delta P_{i \rightarrow f}}{\delta V_{\text{cont}}(x,t)} \nonumber \\
& =  2 \text{Re} \left [ p^{*}(x,t) \psi(x,t) \right ] \\
\label{eq:delL_g_result_final}
\frac{\delta J}{\delta g_{\text{cont}}(x,t)} & \equiv \frac{\delta P_{i \rightarrow f}}{\delta g_{\text{cont}}(x,t)}  \nonumber \\
& = 2 \text{Re} \left [ p^{*}(x,t) |\psi(x,t)|^2 \psi(x,t) \right ],
\end{align}
which must also finally be zero when optimal controls are found. Importantly, eqs. (\ref{eq:delL_V_result_final}) and (\ref{eq:delL_g_result_final}) will generally be nonzero on the way to maximizing $P_{0 \rightarrow f}$.

The evolution of $p(x,t)$ is often referred to as the sensitivity equation, as its spatio-temporal variation and magnitude correspond the the important portions of $\psi(x,t)$ in the controlled dynamics. In the context of the GPE, this can be identified as a mean-field analogue to the Bogoliubov de Gennes (BdG) equation describing the evolution of elementary excitations of the condensate \cite{Pethick_monograph}. Numerically calculating the gradient in eqs. (\ref{eq:delL_V_result_final}) and (\ref{eq:delL_g_result_final}) involves solving eqs. (\ref{eq:dL_dlambda_result_final})-(\ref{eq:delL_delpsi_init}), with special note that eq. (\ref{eq:delL_delpsi_results}) calls for simultaneously solving for $p^*(x,t)$.  Defining the supervector $\vec{p}(x,t) = [p(x,t),p^*(x,t)]^{T}$, then the system of equations can be written as
\begin{align}
\label{eq:p_prop}
\frac{\partial}{\partial t} \vec{p} &= -\frac{i}{\hbar}L \vec{p} \\
L &= \left ( \begin{array}{c c}
[H_0+ V + 2g|\psi|^2] & g\psi^2 \\
-g\psi^{*2} & -[H_0 + V + 2g|\psi|^2] \\ \end{array} \right ).
\end{align}

\section{Propagation methods}
\label{sec:becprop}
Simulation of the GPE requires consideration of numerical stability issues, and much work has been done in this regard with numerous propagation schemes, including semi-implicit Crank-Nicholson methods, time-splitting approaches, and also adaptive step size techniques \cite{Bao_compphyscomm_2013,Chin_pre_2007, Bao_jcompphys_2002, Hohenester_compphyscomm_2014, Borzi_siam_2008, Bao_jcompphys_2003, Muruganandam_compphyscom_2009}. As utilized in ref. \cite{Hohenester_pra_2007}, here we adopt a small time stepping approach based upon symmetric split-order propagation. In this fashion $H_0$ is expressed in momentum space and the potential terms in coordinate space, where a fast Fourier transform $\mathcal{F}_x[\cdot]$ is used to transform between the two representations \cite{Kosloff_jpc_1988}. With suitable resolution over time and space, split-step GPE solutions are known to be of good accuracy \cite{Thalhammer:2009:HTH:1480240.1480283,Bao_jcompphys_2003,Chin_pre_2007}, and a standard second-order symmetric form is employed here (upon setting $\hbar=1$),
\begin{align}
\label{eq:splitting}
&\psi(t+\Delta t) = \nonumber \\
&e^{-i \tilde{H}_1(t+\Delta t)\frac{\Delta t}{2}}e^{-i H_0 \Delta t}e^{-i H_1(t+ \Delta t) \frac{\Delta t}{2}} \psi(t),
\end{align}
where 
\begin{align}
\label{eq:splitting_H1}
H_1(t+\Delta t) & = V(t+\Delta t) + g(t+\Delta t)|\psi(t)|^2 \\
\tilde{H}_1(t+\Delta t) & = V(t+\Delta t) + g(t+\Delta t)|\psi(t+\Delta t)|^2 \\
H_0 &= \hat{p}^2/2m.
\end{align}
Note that in order to calculate the nonlinear term $g |\psi(t)|^2$  in the first step of the splitting procedure, $\psi(x,t)$ from the previous time step is used. The final step in the coordinate representation is performed with increased accuracy by calculating the nonlinear term using $\psi(t+\Delta t)$ at the updated time step. As the final step of the propagation utilizing $\exp(-i \tilde{H}_1(t+\Delta t)\Delta t/2)$ occurs in the coordinate representation, it only adds a phase to $\psi$, allowing for calculation of the modulus $|\psi(t+\Delta t)|$ from the previous two operations
\begin{align}
\label{eq:bec_nlterm_bettersplit}
|\psi(t+\Delta t)| = \left | e^{-i H_0 \Delta t}e^{-i H_1(t+ \Delta t) \frac{\Delta t}{2}}  \psi(t) \right |.
\end{align}
Equation (\ref{eq:bec_nlterm_bettersplit}) can then be used to accurately calculate the nonlinear term in $\tilde{H}_1$ at the updated time step \cite{Hohenester_pra_2007}.

Similar to $\psi$, the coupled system of equations for $\vec{p}$ in eq. (\ref{eq:p_prop}) can be solved through a short time step evolution procedure, which entails backward evolution in time from the terminal condition in eq. (\ref{eq:delL_delpsi_init}),
\begin{align}
\label{eq:p_propagator}
\vec{p}(t-\Delta t) &= \exp\left (i L(t) \Delta t \right) \vec{p}(t).
\end{align}
The calculation of $\exp(i L \Delta t)$ also employs a symmetric splitting scheme, but with special treatment of the potential terms that are no longer diagonal in coordinate space in this coupled equation form: 
\begin{align}
\label{eq:p_prop_split}
&\exp\left (i L(t) \Delta t \right) = e^{i L_1(t) \frac{\Delta t}{2}}e^{i L_0\Delta t}e^{i L_1(t) \frac{\Delta t}{2}} \\
&L_0 = \left (\begin{array}{c c}
H_0 & 0\\
0 & -H_0 \\ \end{array} \right ) \\
&L_1(t) = \left (\begin{array}{c c}
[V + 2g|\psi|^2] & g\psi^2 \\
-g\psi^{*2} & -[ V + 2g|\psi|^2] \\ \end{array}  \right ).
\end{align}
Propagation due to the $L_0$ term is performed in momentum space by Fourier transforming $\vec{p}(x,t)$ to $\vec{p}(k,t)$ as
\begin{align}
\label{eq:ft_pvec}
\vec{p}(k,t) = \mathcal{F}_x[\vec{p}(x,t)] &= \left [ \begin{array}{c}
\mathcal{F}_x[p(x,t)]\\
\mathcal{F}_x[p^*(x,t)]\\ \end{array} \right ],
\end{align}
where $\exp(i L_0 \Delta t)$ is easily calculated, since the $L_0$ term is diagonal in momentum space. In coordinate space the $ L_1$ term at each discrete spatial point $L_1(x_j,t)$ can be represented as a $2\times2$ matrix operator acting on a length two vector $[p(x_j,t), p^*(x_j,t)]^T$, and the matrix exponential $\exp(i L(x_j,t) \Delta t/2)$ of the non-normal matrix $L$ can be expressed as \cite{matrix_anal, Borzi_siam_2008}
\begin{align}
\label{eq:twolevel_prop}
&\exp\left (i L_1(x_j,t) \frac{ \Delta t}{2} \right) = \nonumber \\
&\cos \left(\frac{\Delta t}{2} \lambda_L \right) I + i\sin \left(\frac{\Delta t}{2} \lambda_L \right)\frac{L_1(x_j,t)}{\lambda_L}\\
&\lambda_L = \left ( \left |V(x_j,t)+2g(x_j,t)|\psi(x_j,t)|^2 \right |^2 \right. \nonumber \\
&\quad \quad \left.  - \left | g(x_j,t) \psi^2(x_j,t)\right |^2 \right )^{1/2}.
\end{align}

\section{Nonlinear coherent modes: Identification and Characterization}
\label{sec:bectopomodes}
Commonly examined nonstationary modes of nonlinear Schr\"odinger equations take soliton-like solutions \cite{ablowitz_monograph, debnath_nl_monograph}, but nonlinear coherent modes form  a separate class of stationary modes. Each coherent mode corresponds to a spatial pattern over the entire condensate, making these modes relevant to atom laser technology and the study of large scale quantum relaxation processes \cite{PhysRevA.66.043602, PhysRevA.56.4845, Ramos_pra_2008}. The transitions considered in this work for $P_{0 \rightarrow f}$ optimizations are between coherent modes obtained when $V_{cont}=0$ and $g_{cont}=0$.

Without an analytical form for the coherent modes, they must be numerically determined, which is performed here through an iterative procedure. A technique for determining the ground mode for a time-independent GPE Hamiltonian has been identified based on an imaginary time propagation method, in which the Schr\"odinger equation is evolved in negative imaginary time, such that the amplitude of higher eigenmodes are damped out via factors dependent on their respective energies \cite{Bao_siam_2004,Hohenester_pra_2007}. By a renormalization of the wavefunction to its original norm at each propagation step, the propagation will converge to the lowest-energy eigenmode of the GPE Hamiltonian. Long time propagation of the GPE starting with a purportedly discovered stationary mode in real time provides the means to test if the mode is truly stationary.

\subsection{Spectrum-adapted imaginary time propagation \mbox{(S-AITP)}}
\label{sec:bec_artificial}
The nonlinear coherent modes of interest  $\{\phi_j\}$ are eigenmodes of the nonlinear Hamiltonian operator $H_{\mathrm{nl}}[\phi_j]$,
\begin{align}
\label{eq:bec_ham_gen_nonlinear}
H_{\mathrm{nl}}[\phi_j] \phi_j &=  \left [ H_0 +  \frac{m\omega^2}{2} \left (x-\frac{L}{2} \right)^2+ g_0 |\phi_j|^2 \right ] \phi_j \nonumber \\
&= E_j \phi_j.
\end{align}
The nonlinear Hamiltonian above depends explicitly upon the mode $\phi_j$ on which it operates, and the ground coherent mode $\phi_0$ would be associated with the lowest-energy eigenvalue of $H_{\mathrm{nl}}[\phi_0]$. Locating excited coherent modes is a demanding task, as the conventional imaginary time propagation algorithm will naturally converge to the ground eigenmode of a Hamiltonian. Moreover, simply diagonalizing the Hamiltonian $H_{nl}[\phi_0]$ that can be constructed after convergence of the algorithm to obtain $\phi_0(x)$ does not provide exited modes to the nonlinear Hamiltonian in eq. (\ref{eq:bec_ham_gen_nonlinear}); instead, the latter procedure provides states $\Phi_k^0$ satisfying
\begin{align}
\label{eq:bec_ham_notreallylinear}
H_{\mathrm{nl}}[\phi_0] \Phi_k^0 &=  \left [ H_0 + V_{\text{trap}}(x) + g_0 |\phi_0|^2 \right ] \Phi_k^0 \nonumber \\
&= \mathcal{E}_k^0 \Phi_k^0.
\end{align}
where $\mathcal{E}_k^0$ corresponds to the energy of the linear, Hermitian Hamiltonian $H_{\mathrm{nl}}[\phi_0]$ in eq. (\ref{eq:bec_ham_notreallylinear}). Here, the state $ \Phi_k^0$ is the $k$th eigenstate arising from $H_{\mathrm{nl}}[\phi_0]$, using the $0$-th coherent mode of the desired nonlinear Hamiltonian. Thus, these eigenstates $\Phi_k^0$ are a set of orthonormal states for a Hamiltonian with a potential term utilizing the ground mode $\phi_0$ of the GPE; an analogous  set of separately orthogonal functions $\Phi_k^j$ would arise from the Hamiltonian $H_{\mathrm{nl}}[\phi_j]$. The distinction between the modes $\{\phi_i\}$ and $\{\Phi_k^j\}$ is important, as the latter are not generally stationary modes of the GPE. 

An initial attempt at using projection methods for locating excited modes, which removes contributions from the lower-energy modes at every propagation step, tended to show poor performance (not shown). Alternatively a reverse imaginary time propagation method was developed in which the eigenvalue spectrum of the Hamiltonian is artificially adapted by replacing the smallest eigenvalue with a very large one, such that the lowest-lying eigenmode is always dampened more than other excited modes during execution of the algorithm. In this manner, any desired excited mode can be self-consistently incorporated into the Hamiltonian. This scheme is referred to here as spectrum-adapted imaginary time propagation (S-AITP), and it is capable of locating arbitrary order excited modes. The basic schematic of the algorithm is given below for identifying the $j$-th eigenmode $\phi_j$. For every time step in which imaginary time propagation would conventionally be calculated as $\psi_j(t-i\Delta t) = \exp(-H_{\mathrm{nl}}(t)\Delta t) \psi_j(t)$, the following procedure is instead performed:

\begin{itemize}
\item{Diagonalize the Hamiltonian $H_{\mathrm{nl}}[\psi_j(t)](t) = H_0 + V + g_0|\psi_j(t)|^2$ into a matrix of eigenvalues $\Lambda = \text{diag}(\lambda_1,\lambda_2,...\lambda_N )$ and associated eigenvectors $M = [f_1, f_2,...f_N]$, where $\lambda_1 < \lambda_2 < ... \lambda_N$. The wavefunction used to generate the nonlinear term will the be the most updated wavefunction from the previous iteration, hence the explicitly labeled time-dependence of the Hamiltonian. }
\item{Create the adapted spectrum $\tilde{\Lambda}$ by replacing the smallest $j-1$ eigenvalues with  $\lambda_N$ in order to locate state $\phi_j$. \mbox{$\tilde{\Lambda} =  \text{diag}(\lambda_N,\lambda_N,..\lambda_N,\lambda_j, \lambda_{j+1},....\lambda_N )$.}}
\item{Calculate the propagation term as $\tilde{U} = M^{\dag}\text{exp}(-\tilde{\Lambda} \Delta t)M$.}
\item{Propagate $\psi_j(x,t-i\Delta t) = \tilde{U}\psi_j(x,t)$.}
\item{Renormalize $\psi_j(x,t-i\Delta t)$ to have norm 1.}
\item{Repeat for $k$ iterations until the norm of the distance \\ \mbox{$\| \psi_j(x,t-ki\Delta t)-\psi_j(x,t-[k-1]i\Delta t)\|^2$} is below a specified threshold value $\epsilon$, indicating that the eigenmode is adequately identified as \mbox{$\psi_j(x,t-ki\Delta t) = f_j \equiv \phi_j$}, modulo a phase factor.}
\end{itemize}

As this is a self-consistent, iterative method, the initial trial wavefunction $\psi_j(t=0)$ can be important for efficiently locating a stationary excited mode. Diffusion Monte Carlo is a related technique utilized in electronic structure calculations, and under that technique trial states can successfully converge regardless of their relative energies, provided that the initial guess maintains the correct spatial wavefunction symmetry about the center of the trapping potential (\emph{i.e.}, even or odd parity wavefunctions) \cite{PhysRevB.60.4558}. This suggests that similar initial choices in imaginary time propagation can also be effective at locating excited modes. For example, ref. \cite{Bao_siam_2004} reported success at locating the first excited mode of a BEC in a harmonic trap with conventional imaginary propagation techniques by choosing an odd-parity spatial function as the initial trial guess. While locating an excited mode in systems containing an asymmetric trapping potential falls outside of the scope of this strategy for choosing an initial wavefunction guess, additional considerations about the symmetry group of the irreducible representation of the Hamiltonian can provide similar guarantees for wavefunction convergence in diffusion Monte Carlo, which may also extend to the imaginary time propagation search techniques discussed here \cite{PhysRevB.60.4558}.  In the following section, higher-order coherent modes in a symmetric trapping potential are located with the S-AITP method.

\subsection{Nonlinear coherent modes of a BEC in a harmonic trap}
\begin{figure*}[htb]
\begin{center}
\subfloat{\includegraphics[width=0.3\textwidth, height=0.3\textheight, keepaspectratio]{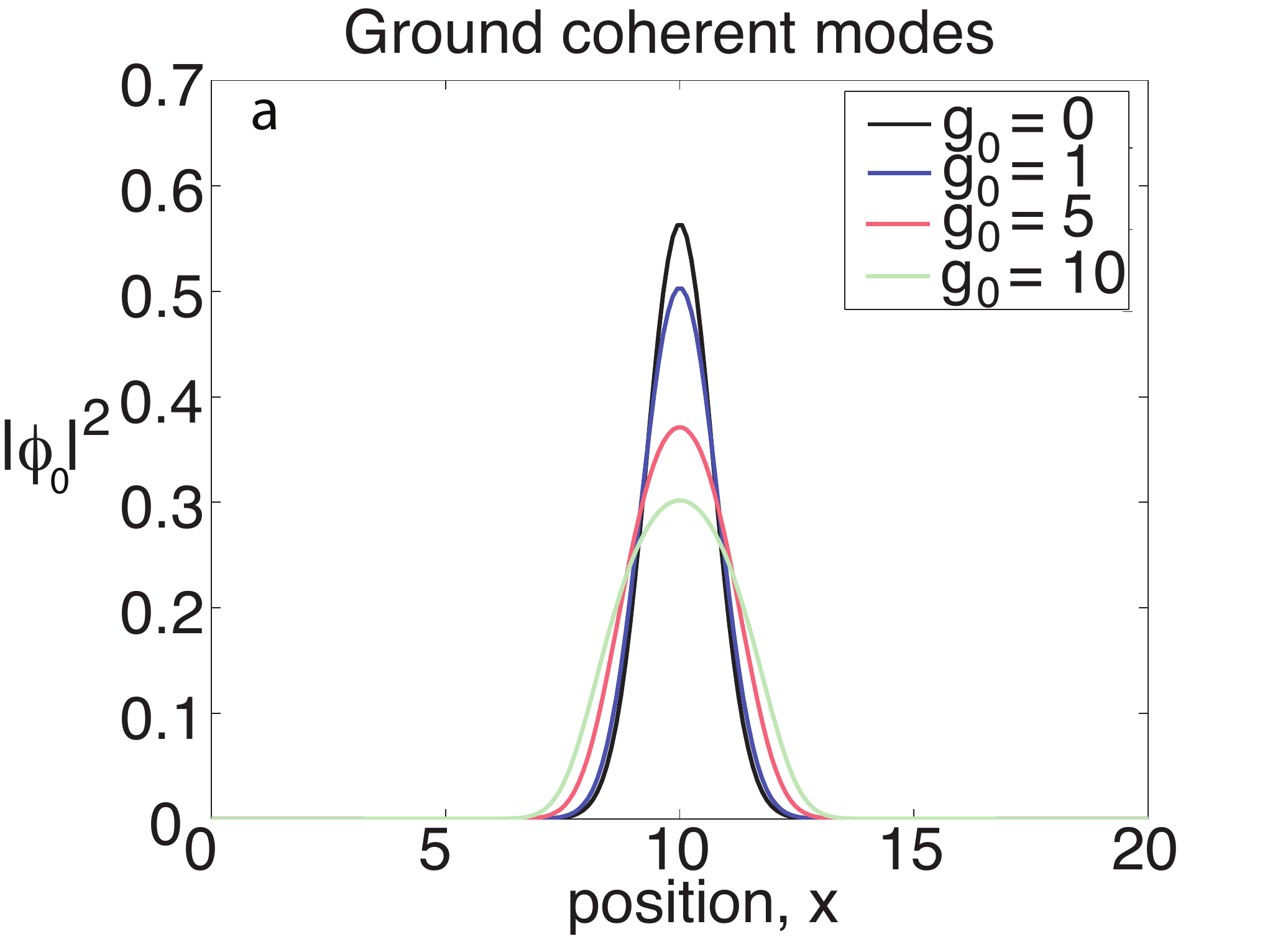}}
\subfloat{\includegraphics[width=0.3\textwidth, height=0.3\textheight, keepaspectratio]{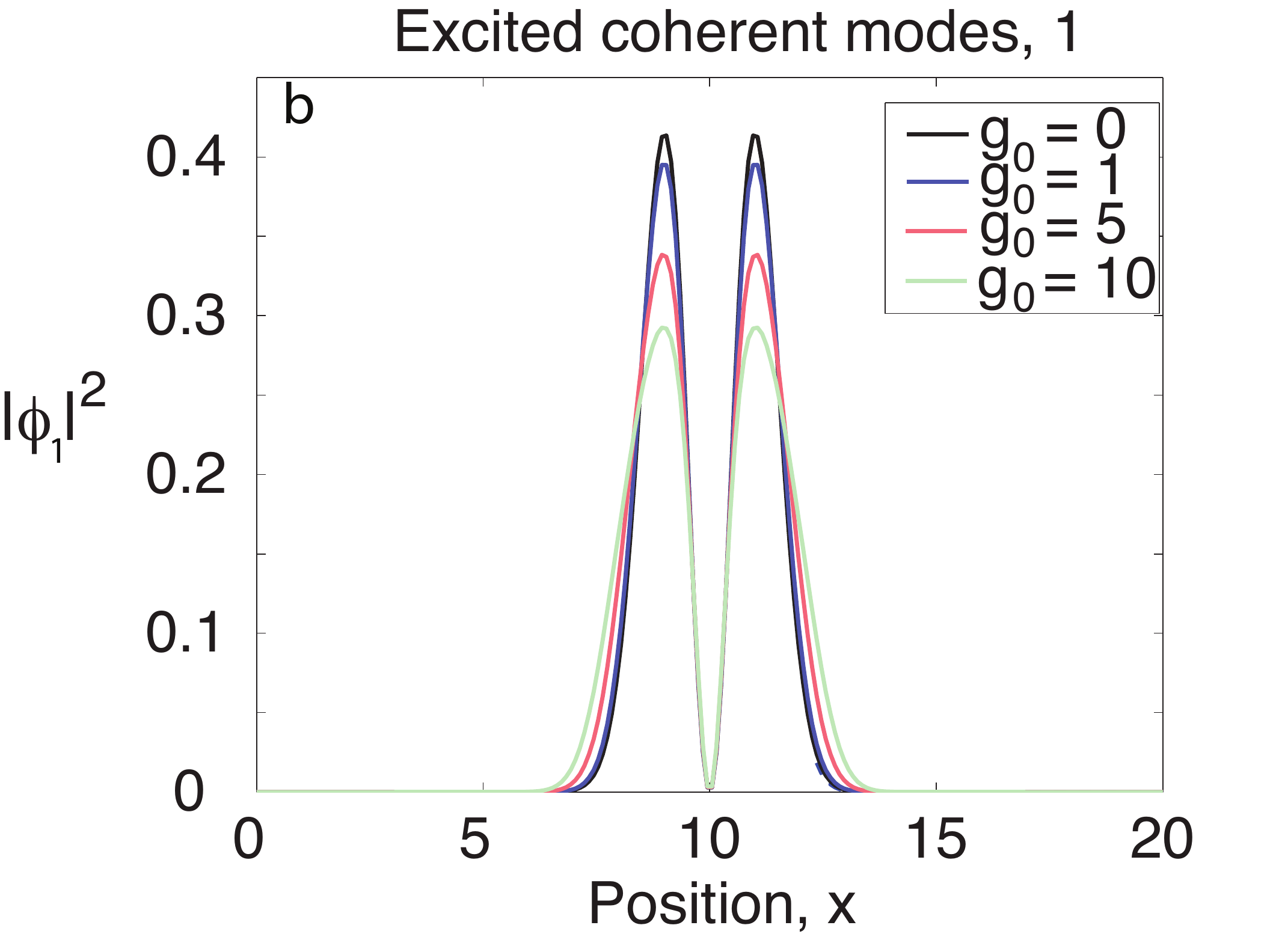}}
\subfloat{\includegraphics[width=0.3\textwidth, height=0.3\textheight, keepaspectratio]{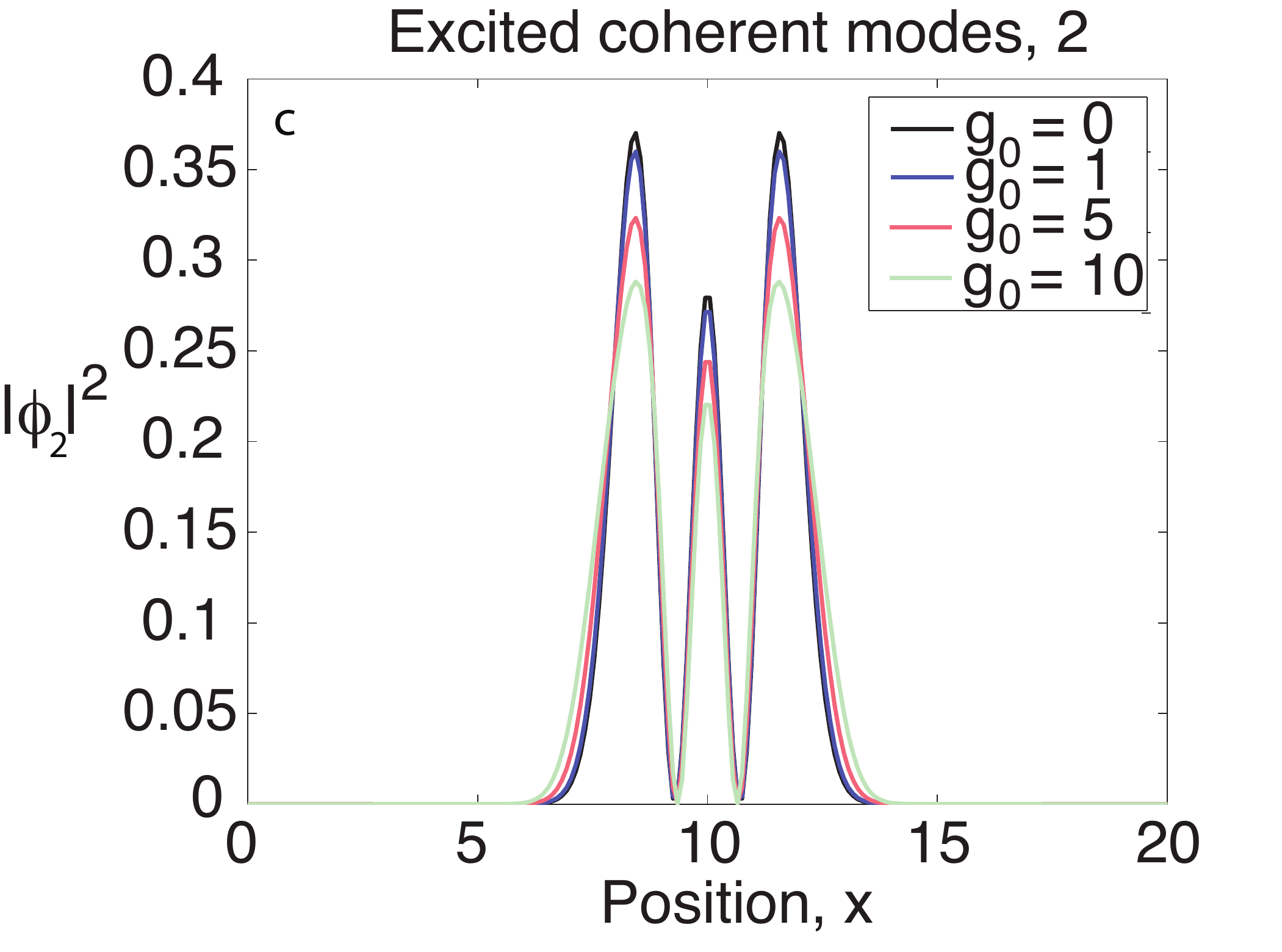}}
\end{center}
\begin{center}
\subfloat{\includegraphics[width=0.315\textwidth, height=0.315\textheight, keepaspectratio]{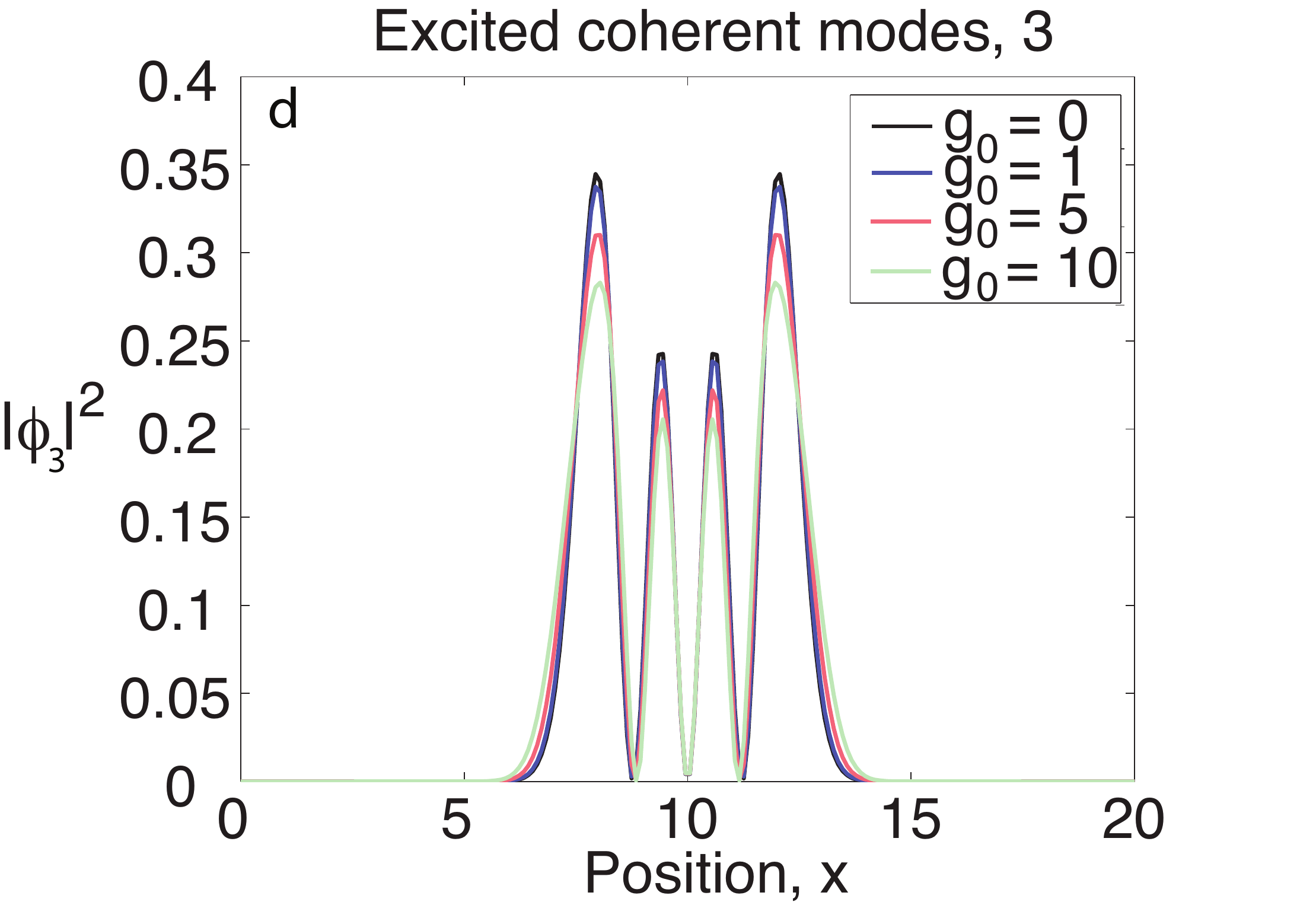}}
\subfloat{\includegraphics[width=0.3\textwidth, height=0.3\textheight, keepaspectratio]{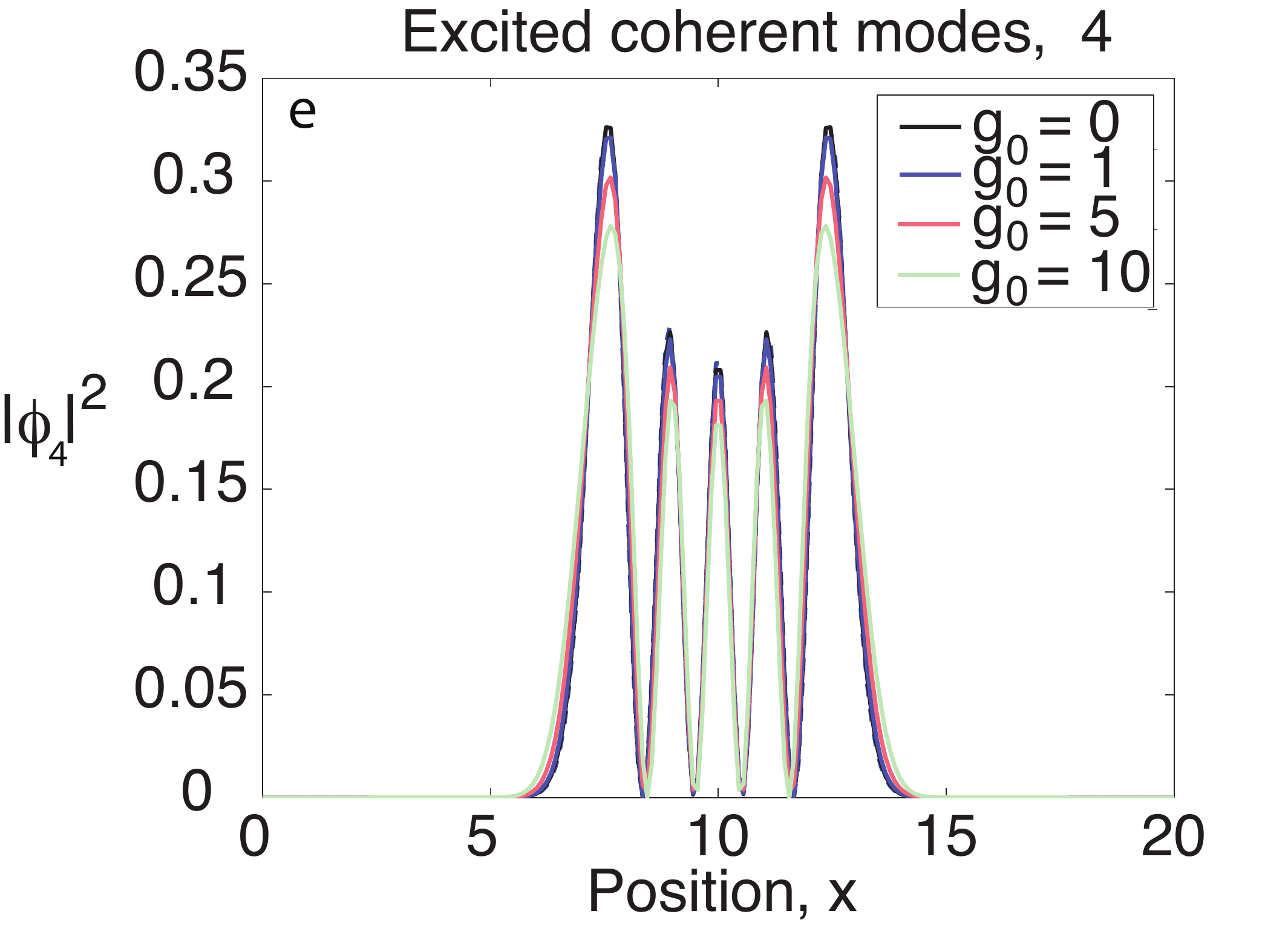}}
\subfloat{\includegraphics[width=0.3\textwidth, height=0.3\textheight, keepaspectratio]{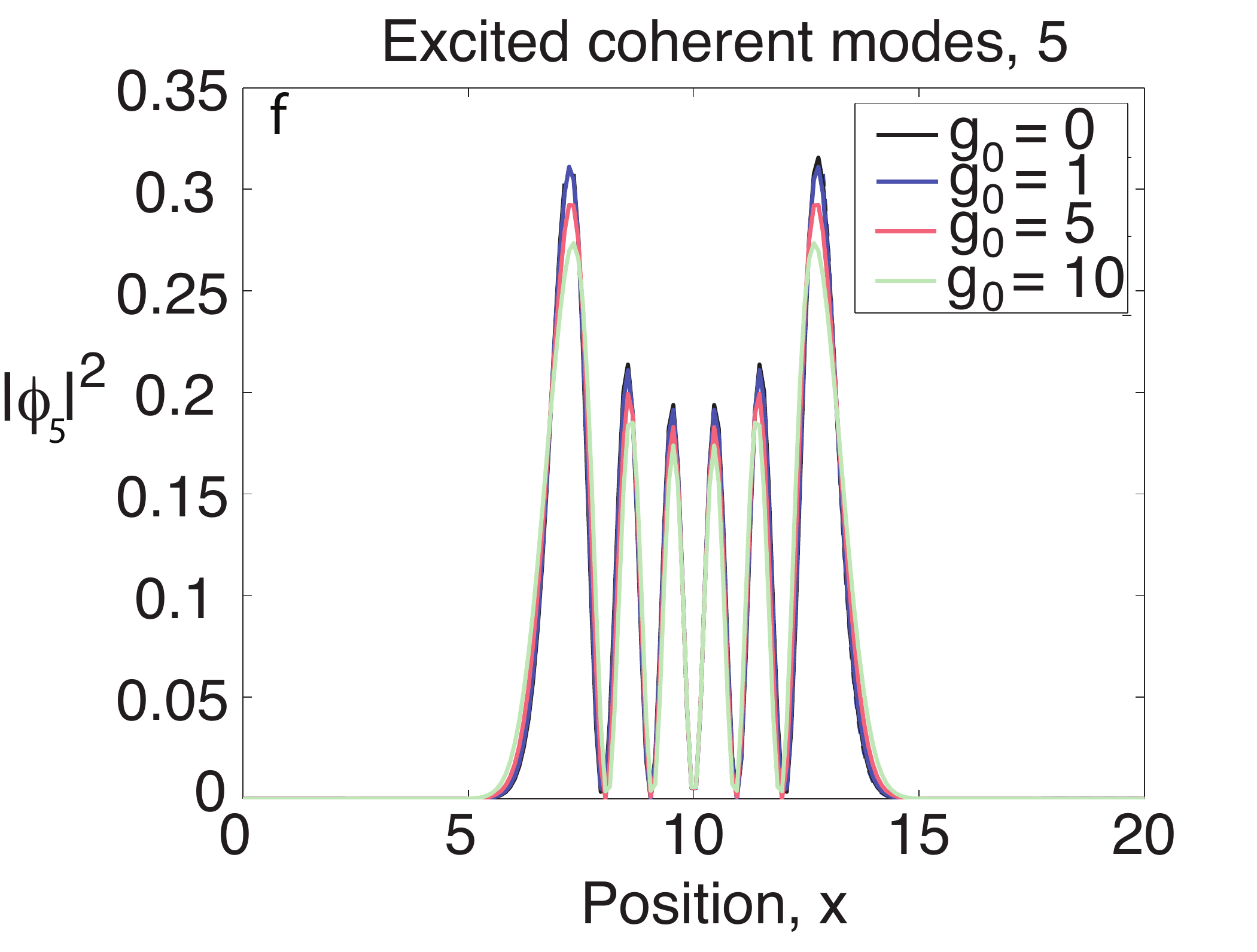}}
\end{center} 
\caption{\label{fig:state0_1_2} Coherent modes  densities $|\phi_j|^2$, $j = 0,1,...5$ for various $g_0$ values.}
\end{figure*}

The GPE's nonlinear contribution does not change the symmetry of the Hamiltonian, nor the parity of its coherent excited modes in accord with the linear case ($g_0=0$). Thus, the linear quantum harmonic oscillator Hamiltonian wavefunctions are chosen as suitable initial trials to the \mbox{S-AITP} for determining the coherent stationary modes of the GPE. Using linear-case wavefunctions as trial forms for S-AITP and a convergence criteria of $\epsilon < 10^{-10}$, the first five excited coherent modes of a control-free BEC (\emph{i.e.}, eq. (\ref{eq:bec_ham_gen_nonlinear})) with varying nonlinearity strengths $g_0$ were located.  The length of the spatial grid is $L = 20$ ($\Delta x = L/300$). To examine nonlinearity strengths in the regime from weak up to modest strength (compared to the trap frequency $\omega$), various values of $g_0 \in [0,1,5,10,20]$ were utilized.  The square amplitudes $|\phi_j|^2$ of the located coherent modes for the ground and first five excited modes are shown in Figure \ref{fig:state0_1_2}. Comparing the different excited modes to the linear Hamiltonian states ($g_0=0$) shows that the $j$-th harmonic oscillator state is a reasonable initial guess to the $j$-th GPE coherent mode, as the higher excited GPE stationary modes appear very similar in form across a range of nonlinearity strengths $g_0 > 0$. This is perhaps due to a reduced contribution from the nonlinear term in the Hamiltonian when the local wavefunction density is lower in these higher energy modes (See the ordinate scales in Figure \ref{fig:state0_1_2}).

A distance measure $d(\phi_j)$ was used to assess the stability of solutions of the coherent modes $\{\phi_j\}$:
\begin{align}
\label{eq:stability_measure}
d(\phi_j) &= 1-\left| \int_0^L \phi_j^*(x) \psi(x,T) dx \right |^2 \\
&\quad \psi(x,0) = \phi_j(x),
\end{align}
where $\psi(x,T)$ is the result of propagating $\psi(x,0)= \phi_j(x)$ forward in real time. The distance measures $d(\phi_j)$ for different $g_0$ values are listed in Table \ref{tab:distance} for propagation time $T=10$ ($\Delta t = T/500$). While the stability of higher energy modes is diminished, all of the identified stationary modes accumulate less than $0.1\%$ error. Reducing the convergence criteria $\epsilon$ and refining the spatial grid can increase the stability of the stationary modes, although we found the present performance suitable for defining an optimization target.

An important aspect for understanding transitions among nonlinear coherent modes is their energy level structure for different nonlinearity strengths $g_0$,
\begin{align}
\label{eq:becenergy}
E_j &= \langle \phi_j |H_{\mathrm{nl}}[\phi_j] | \phi_j \rangle \nonumber \\
&= \int_0^L dx  \phi_j^*(x) \left [ -\frac{1}{2m}\frac{d^2}{d x^2} +   \frac{m \omega^2}{2}\left(x-\frac{L}{2}\right)^2 \right .  \nonumber \\
&\quad  \quad \quad \quad +  \left . \vphantom{\frac{\nabla^2}{2m}} g_0|\phi_j(x)|^2 \right ] \phi_j(x),
\end{align}
where the second derivative is approximated by a fourth-order finite difference expression. The energy level differences $E_f-E_j$, from  Eq. (\ref{eq:becenergy}) will be utilized to characterize the frequency components in the control potentials that promote transitions between the modes. Table \ref{tab:becenergy} lists the energies of the coherent modes for various values of $g_0$. The linear quantum harmonic oscillator at $g_0=0$ has constant energy level spacings between successive modes (states), and this transition energy degeneracy is broken for GPE nonlinear coherent modes. The originally degenerate level spacings grow further apart for successive higher-energy modes for any value of $g_0$, suggesting an anharmonic ``steeping" contribution to the harmonic trap from the nonlinear term. The effect may be qualitatively understood from the outer two strong, positive features in $|\phi_j(x)|^2$, evident in Figure \ref{fig:state0_1_2}.

\begin{table}[htb]
 \caption{\label{tab:distance} Stability distance measure $d(\phi_j)$ for nonlinear coherent modes $\phi_j$ in eq. (\ref{eq:stability_measure}). }
\begin{center}
\begin{tabular}{l|  l l l l l}
 \hline
 $g_0$&   0&1 & 5&  10&  20 \\
\hline
$\phi_0$& $4 \times10^{-10}$&  $2 \times10^{-7}$ & $6 \times10^{-7}$&  $1 \times10^{-6}$&  $2\times10^{-6}$\\
$\phi_1$& $1 \times10^{-9}$&  $3 \times10^{-6}$ & $7 \times10^{-7}$&  $4 \times10^{-6}$&  $1 \times10^{-5}$\\
$\phi_2$& $3 \times10^{-9}$&  $1 \times10^{-5}$ & $1 \times10^{-5}$&  $1 \times10^{-5}$&  $6 \times10^{-5}$\\
$\phi_3$& $5 \times10^{-9}$&  $3 \times10^{-5}$ & $5 \times10^{-5}$&  $5 \times10^{-5}$&  $8 \times10^{-5}$\\
$\phi_4$& $8 \times10^{-9}$&  $6\times10^{-5}$ & $1 \times10^{-4}$&  $2 \times10^{-4}$&  $1 \times10^{-4}$\\
$\phi_5$& $1 \times10^{-8}$&  $1 \times10^{-4}$ & $3 \times10^{-4}$&  $4 \times10^{-4}$&  $3 \times10^{-4}$\\
\hline
\end{tabular}
\end{center}
\end{table}

\begin{table}[htb]
 \caption{\label{tab:becenergy} Calculated energy $E_j$ of nonlinear coherent modes $\phi_j$. }
\begin{center}
\begin{tabular}{l|  l l l l l}
 \hline
 $g_0$&  0&1 & 5&  10&  20 \\
\hline
$E_0$&  0.50 & 0.87 & 2.01 & 3.11  &4.87   \\
$E_1$&  1.50 & 1.79 & 2.81 & 3.86  &5.61  \\
$E_2$&  2.50 & 2.75 & 3.67 &4.68   &6.39   \\
$E_3$&  3.50 &  3.73    & 4.58  &5.54  &7.20   \\
$E_4$&  4.50 & 4.71 &5.51   &6.42  &8.04   \\
$E_5$&  5.50 & 5.69 & 6.45  &7.33 &8.90 \\
\hline
\end{tabular}
\end{center}
\end{table}

\begin{figure}[htb]
\begin{center}
\subfloat{\includegraphics[width=\textwidth, height=\textheight, keepaspectratio]{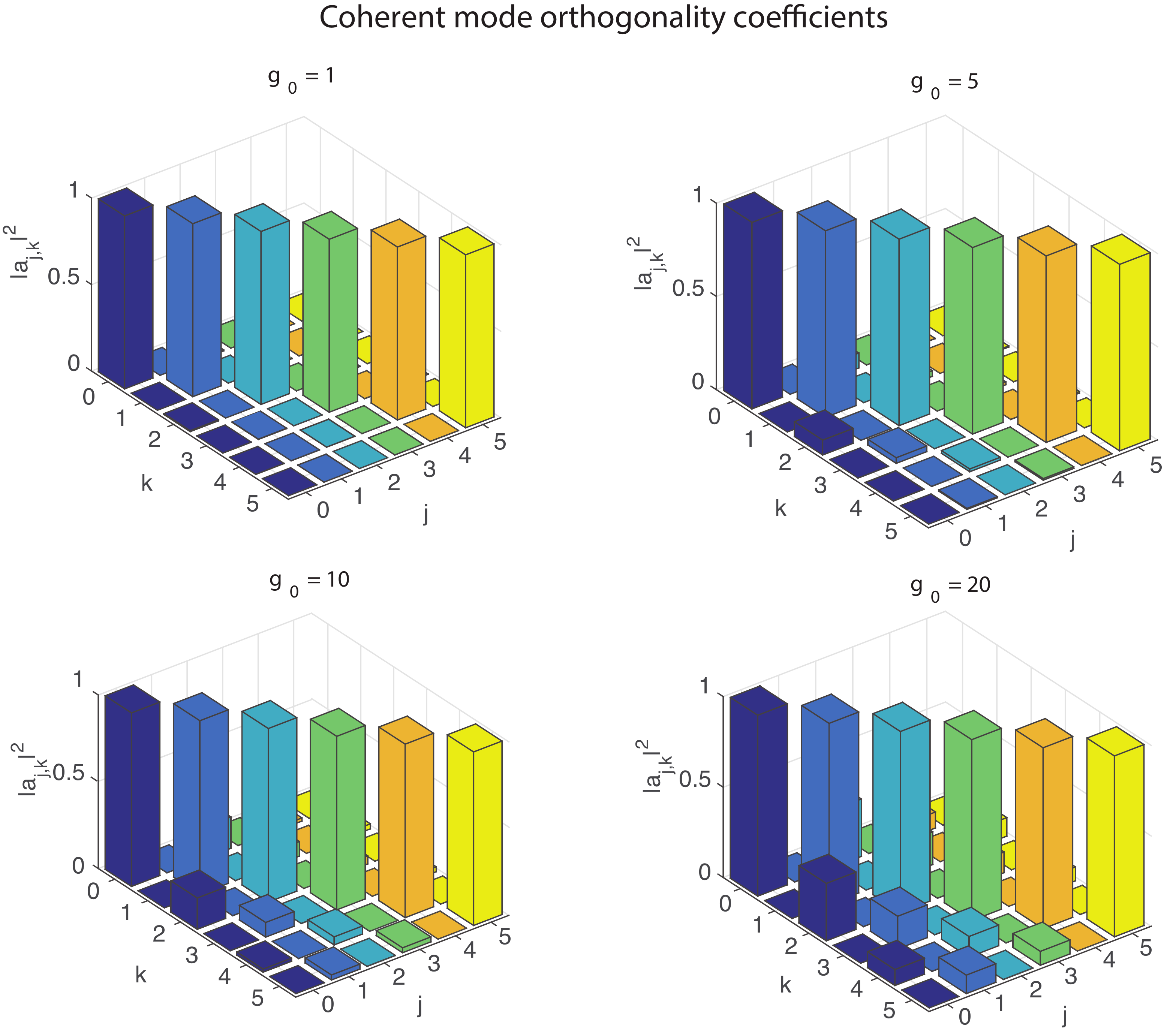}}
\caption{\label{fig:bec_mutualorthog} Orthogonality coefficients $|a_{j,k}|^2 = |\int_0^L \phi_j^*(x) \phi_k(x) dx|^2$. Orthogonality is preserved at small $g_0$, although at larger $g_0$ values the lower-lying modes begin to overlap. Nevertheless, the set of functions $\{\phi_i\}$ show a high degree of orthogonality, despite each arising from corresponding distinct Hamiltonians $H_{\mathrm{nl}}[\phi_i]$.}
\end{center}
\end{figure}

As each individual coherent mode $\phi_j$ arises from its associated Hermitian Hamiltonian $H_{\mathrm{nl}}[\phi_j]$, rather than a single Hamiltonian as in eq. (\ref{eq:bec_ham_notreallylinear}),  the coherent modes are not expected to be orthogonal. The degree of orthogonality between modes $\phi_j(x)$ and $\phi_k(x)$ is captured through the overlap term $a_{j,k}$,
\begin{align}
\label{eq:gpd_modeorthog}
a_{j,k} &= \int_0^L \phi_j^*(x) \phi_k(x) dx .
\end{align}
The square amplitudes of the orthogonality coefficients  $|a_{j,k}|^2$for the stationary BEC modes are given in Figure \ref{fig:bec_mutualorthog} for the each value of $g_0$. For weak nonlinear contributions, the overlaps show that the nonlinear modes demonstrate a high degree of orthogonality, though this diminishes for larger $g_0$ values. Additionally, the reduction in orthogonal behavior is most evident for lower-energy modes, due to the more prominent effect for the nonlinear term in the Hamiltonians in this energy regime.

\section{Optimization of coherent mode-to-mode transitions}
\label{sec:becoct}
Maximization of the mode-to-mode $P_{0\rightarrow f}$ transitions is considered here for a variety of $g_0$ values. Three separate spatio-temporal control schemes are studied: (i)  control by solely the potential $V_{\mathrm{cont}}(x,t)$, (ii) dual control using both $V_{\mathrm{cont}}(x,t)$ and the nonlinear strength $g_{\mathrm{cont}}(x,t)$, and (iii) nonlinearity-only control with just $g_{\mathrm{cont}}(x,t)$. Initial trial controls for the optimization take the form of a simple sinusoidal oscillation in both time and space for the potential control and a constant value for $g_{\mathrm{cont}}$,
\begin{align}
\label{eq:V_init}
V_{\mathrm{cont}}(x,t) &= a \sin(\omega_v t - \alpha(x)), \\
\label{eq:V_init1}
g_{\mathrm{cont}}(x,t) &= g. 
\end{align}
Several initial trials were chosen for $V_{\mathrm{cont}}(x,t)$ in the study, where $a =$ either 1 or $1/5\pi$, as well as spatially independent ($\alpha(x)  = \pi/L$) or spatially dependent ($\alpha(x) = \pi x/L$) phases. Note that in the case $\alpha(x)= \pi x/L$ we have \mbox{$\sin(\omega_v t - \alpha(x))$} = \mbox{$\sin(\omega_v t)\cos(\alpha(x))$} - \mbox{$\cos(\omega_v t)\sin(\alpha(x))$} such that the trial potential has both symmetric and  anti-symmetric spatial components.  In all cases the initial frequency was set to $\omega_v=\omega/10$.  The latter low frequency in the non-resonant trials of eq. (\ref{eq:V_init}) was chosen to initially couple into the system, while minimally overlapping with system transition frequencies so as not to bias the final optimal control (See Sec. \ref{sec:bec_analysis}). The constant nonlinearity trial was set to $g_{\mathrm{cont}}(x,t)=g_0$ for potential-only control. For dual and nonlinearity-only control,  different constant values of $g_{\mathrm{cont}}(x,t) \in \pm [1,5,10,20] $ were all examined as initial trial guesses. Initial $g_{\mathrm{cont}}$ values were chosen from this range such that $(g_\mathrm{cont}+g_0) \in [1,5,10,20]$ for any given trial. This ensured that any trial nonlinearity was always positive. The choice of a fixed ``background potential" $V_{\mathrm{cont}}(x,t)$ being present can significantly impact the effectiveness of $g_{\mathrm{cont}}(x,t)$ as a spatio-temporal control. Thus, for nonlinear-only control, the trial potential form in eq. (\ref{eq:V_init}) was also included (utilizing each form of $\alpha(x)$ and $a$).  Although not functionally varied throughout the optimization, it was found that such a background potential term aided the optimization efficiency for $g_{\mathrm{cont}}(x,t)$. Two terminal times $T=\pi$ and $T=10$ are studied,  with $\Delta t$ = $\pi/500$, and the length of the space domain is $L = 20$, with $\Delta x = L/300$. Optimization is performed with the D-MORPH algorithm \cite{Rothman_pra_2005,Rothman_pra_2006,Moore_pra_2011} using the gradient forms in eq. (\ref{eq:delL_V_result_final}) and (\ref{eq:delL_g_result_final}). Specifically, a generic control $c(x,t) = V_{\mathrm{cont}}(x,t)$ or $ g_{\mathrm{cont}}(x,t)$ depends on the search trajectory variable $s \geq 0$ with the requirement that $dP_{0\rightarrow f}/ds \geq 0$ ,
\begin{align}
\label{eq:becdmorph1}
\frac{d P_{0 \rightarrow f}}{d s} &= \int_0^L\int_0^T \frac{\delta P_{0 \rightarrow f}}{\delta c(s,x,t)}\frac{\partial c(s,x,t)}{\partial s} dt dx \quad \geq 0,
\end{align}
assured by 
\begin{align}
\label{eq:becdmorph2}
\frac{\partial c(s,x,t)}{\partial s} &= \frac{\delta P_{0 \rightarrow f}}{\delta c(s,x,t)}.
\end{align}
Eq. (\ref{eq:becdmorph2}) is numerically solved with a fourth-order Runge-Kutta integrator (MATLAB's \texttt{ode45} routine), where each optimization step by \texttt{ode45} increases the trajectory variable $s$. In the case of dual controls, then two corresponding relations like eq. (\ref{eq:becdmorph2}) were used. Optimizations were stopped once $P_{0 \rightarrow f} > 0.99$. 

\begin{figure}[htb]
\begin{center}
\subfloat{\includegraphics[width=\textwidth, height=\textheight, keepaspectratio]{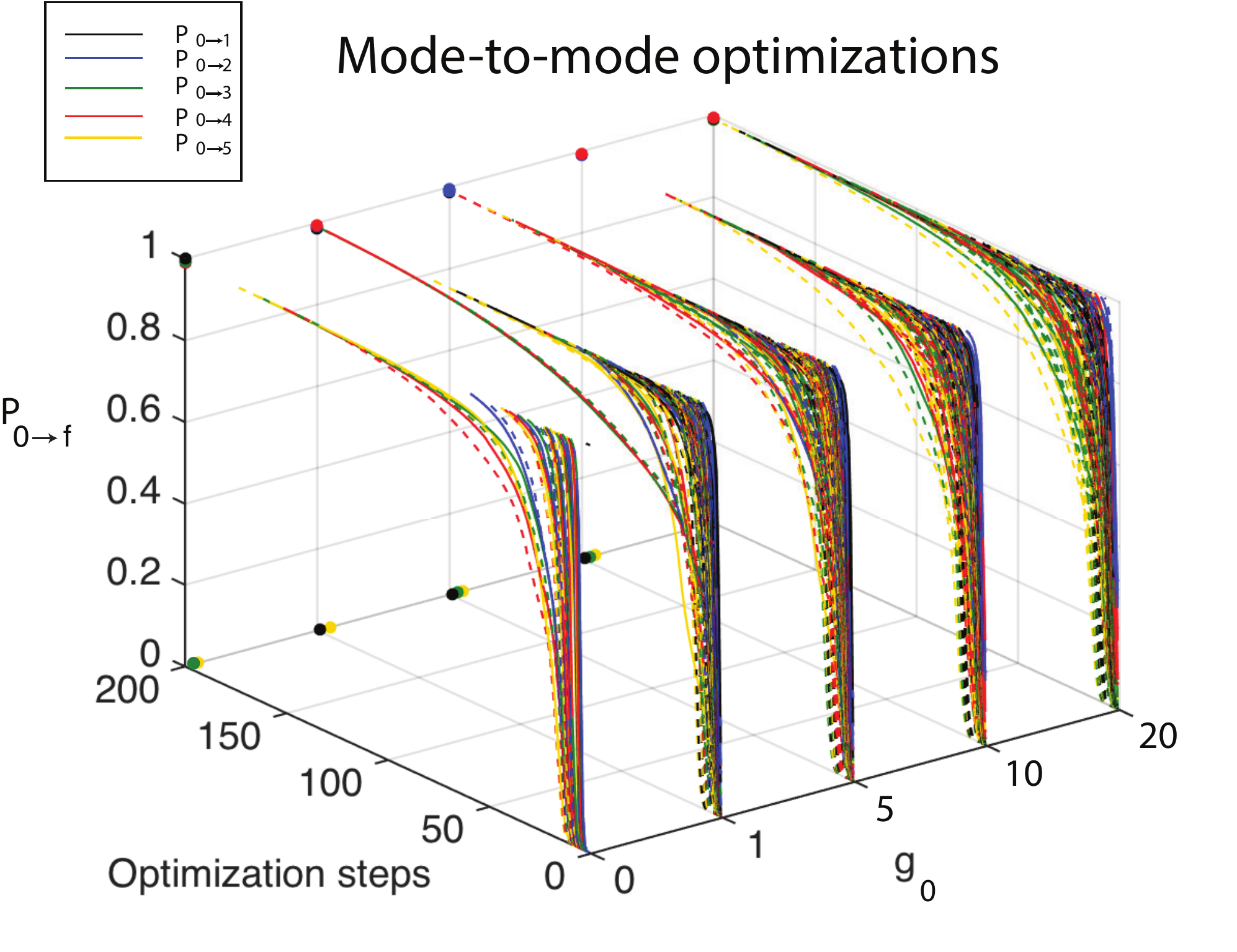}}
\caption{\label{fig:optall} Optimal control trajectories for all initial conditions and nonlinearity strengths. Final trajectory values are projected on the far axis with dots. Dotted lines correspond to spatially symmetric initial guesses (\emph{i.e.}, $\alpha(x) = \pi/L$) to $V_{\mathrm{cont}}(x,t)$, while solid lines are coupled in the spatially dependent phase of eq. (\ref{eq:V_init}) ($\alpha(x)  = \pi x/L$ ). The trajectories that failed to optimize away from  $P_{0 \rightarrow f} = 0$ used the spatially symmetric initial conditions, and could not readily grow in the antisymmetric potential necessary for optimization to a desired antisymmetric objective mode. A small portion of the simulations ($\approx 5\%$) have been omitted for visualization purposes, as they took longer than 200 optimization steps.}
\end{center}
\end{figure}

The optimization trajectories (\emph{i.e.}, $P_{0 \rightarrow f}$ progression over each optimization step) of all initial control and target combinations are shown in Figure \ref{fig:optall}. In total, over 700 optimizations were performed. All trials utilizing initial guesses with a spatially dependent phase $\alpha(x) = \pi x/L$ succeeded with $P_{0\rightarrow f} > 0.99$. A small portion of solutions ($~6\%$) of the spatially-dependent phase guesses initially led to suboptimal fidelities at the given spatial and temporal resolution of the controls; however, upon refining the temporal and spatial resolution all of the controls resulted in optimization $P_{0\rightarrow f} > 0.99$. Spatial and temporal meshing are known to act as ``resources" in the linear Schr\"odinger equation where adequate resolution is needed to avoid false suboptimal solutions \cite{PhysRevA.90.013404}; the findings above are consistent with the latter resource considerations. However, the spatially independent initial guesses displayed different behavior. Roughly a third of the \emph{asymmetric} target trials for dual control were not successful when using \emph{symmetric} initial control guesses with $\alpha(x) = \pi/L$. In this instance of a spatially symmetric control paired with an asymmetric target, the initial fidelity value of a control is always $P_{0\rightarrow f} \sim 0$, and poses a strong challenge for the numerical optimization to grow in asymmetric features from a critical point at the \emph{bottom} of the quantum control landscape. Interestingly, this symmetry mismatch was still still overcome in many instances (\emph{i.e.}, likely due to numerical ``noise" breaking the symmetry). Such a result draws attention to the practical considerations involved when choosing ``reasonable guesses" to a quantum control optimization.

\section{Dynamical control analysis}
\label{sec:bec_analysis} 
This section seeks to examine the qualitative features of the identified controls. When traditionally examining optimal control results from the linear Schr\"odinger equation, the primary features of the controls and population evolution can aid in establishing the dynamical mechanism, and we follow a similar path here for relating GPE dynamics to key features of the controls. Importantly, the structure identified in these controls does not exclude a wide array of other possible optimal controls for a particular objective.

\begin{figure*}[htb]
\begin{center}
\subfloat{\includegraphics[width=0.245\textwidth, height=0.245\textheight, keepaspectratio]{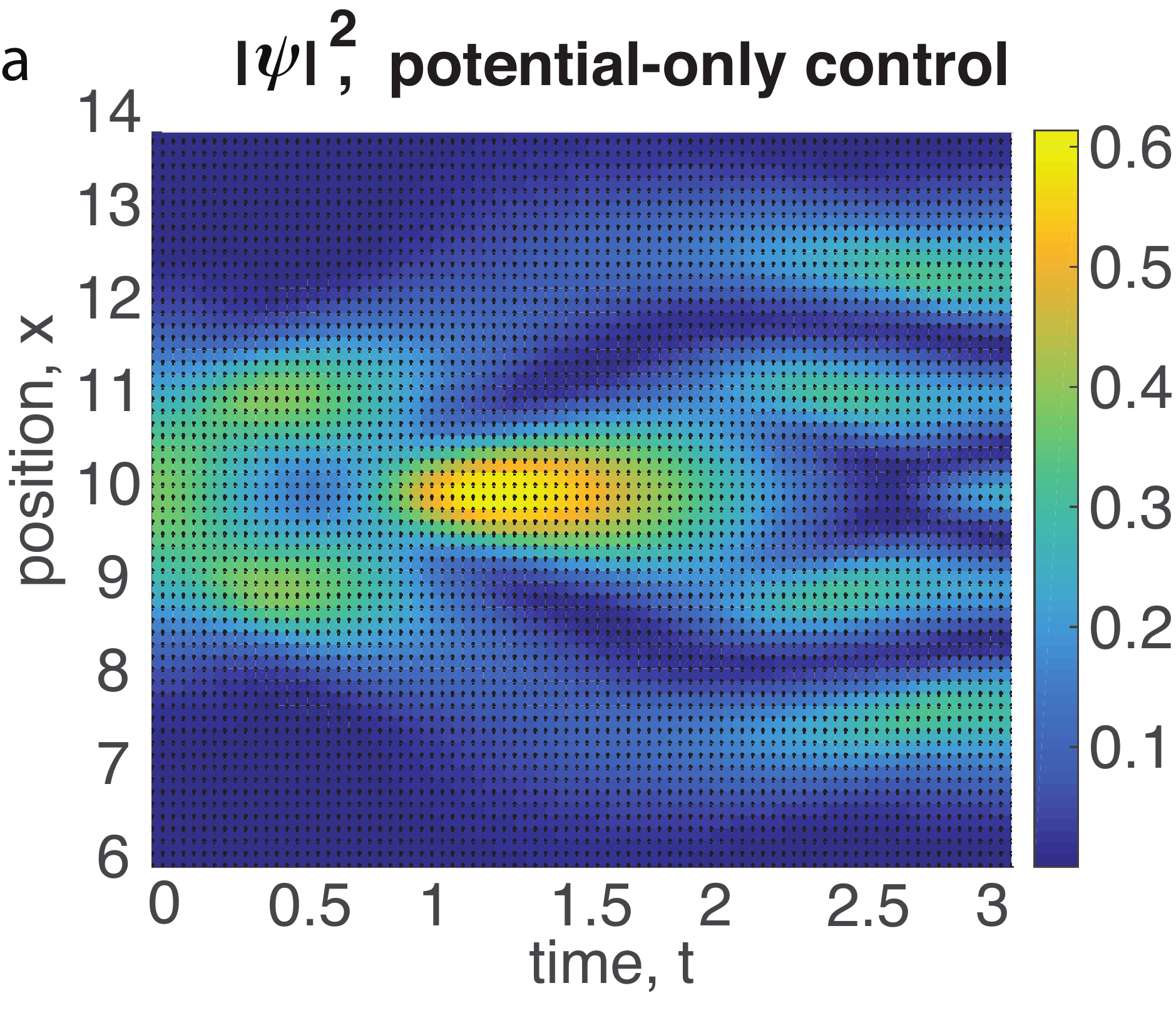}}
\subfloat{\includegraphics[width=0.25\textwidth, height=0.25\textheight, keepaspectratio]{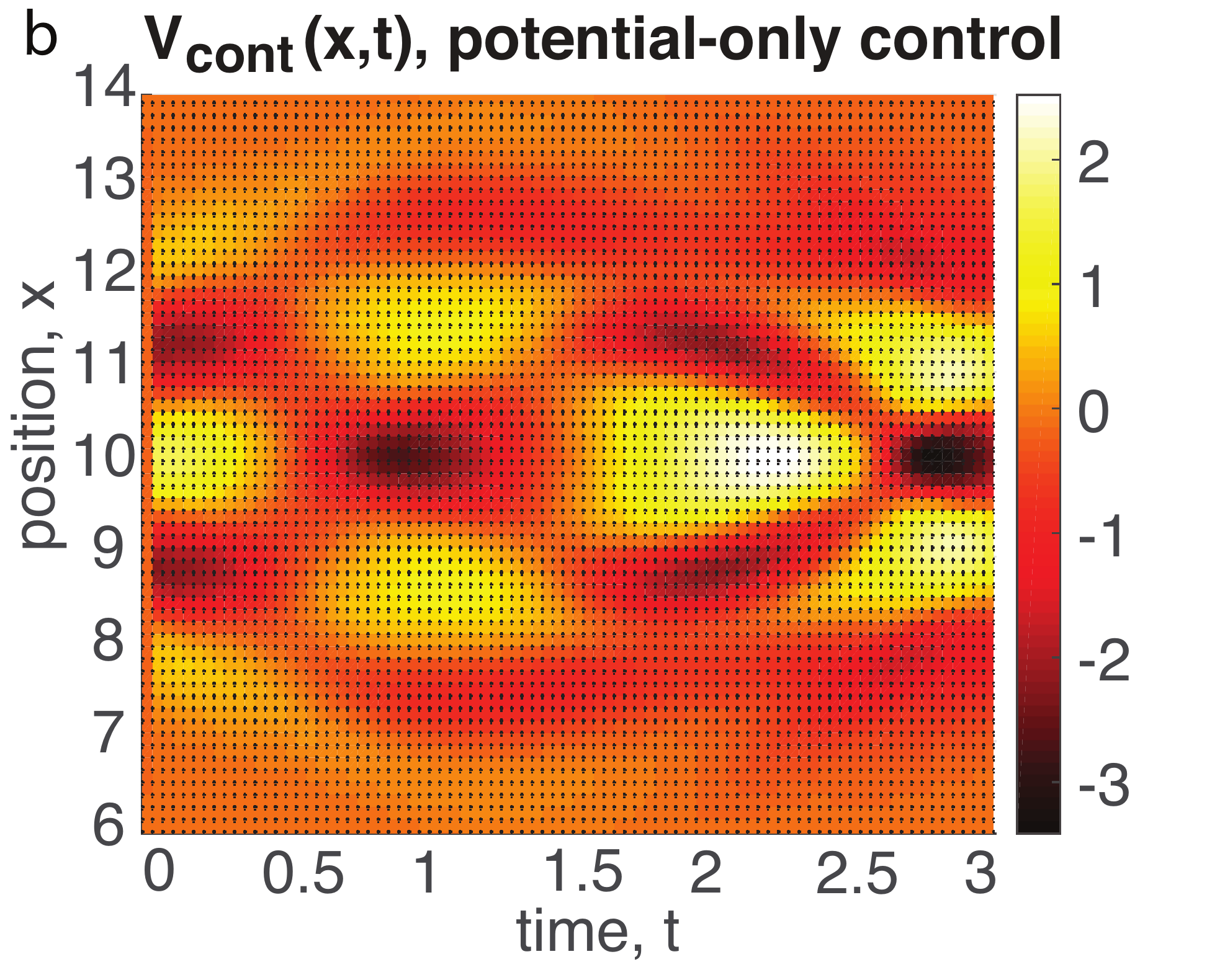}}
\subfloat{\includegraphics[width=0.28\textwidth, height=0.28\textheight, keepaspectratio]{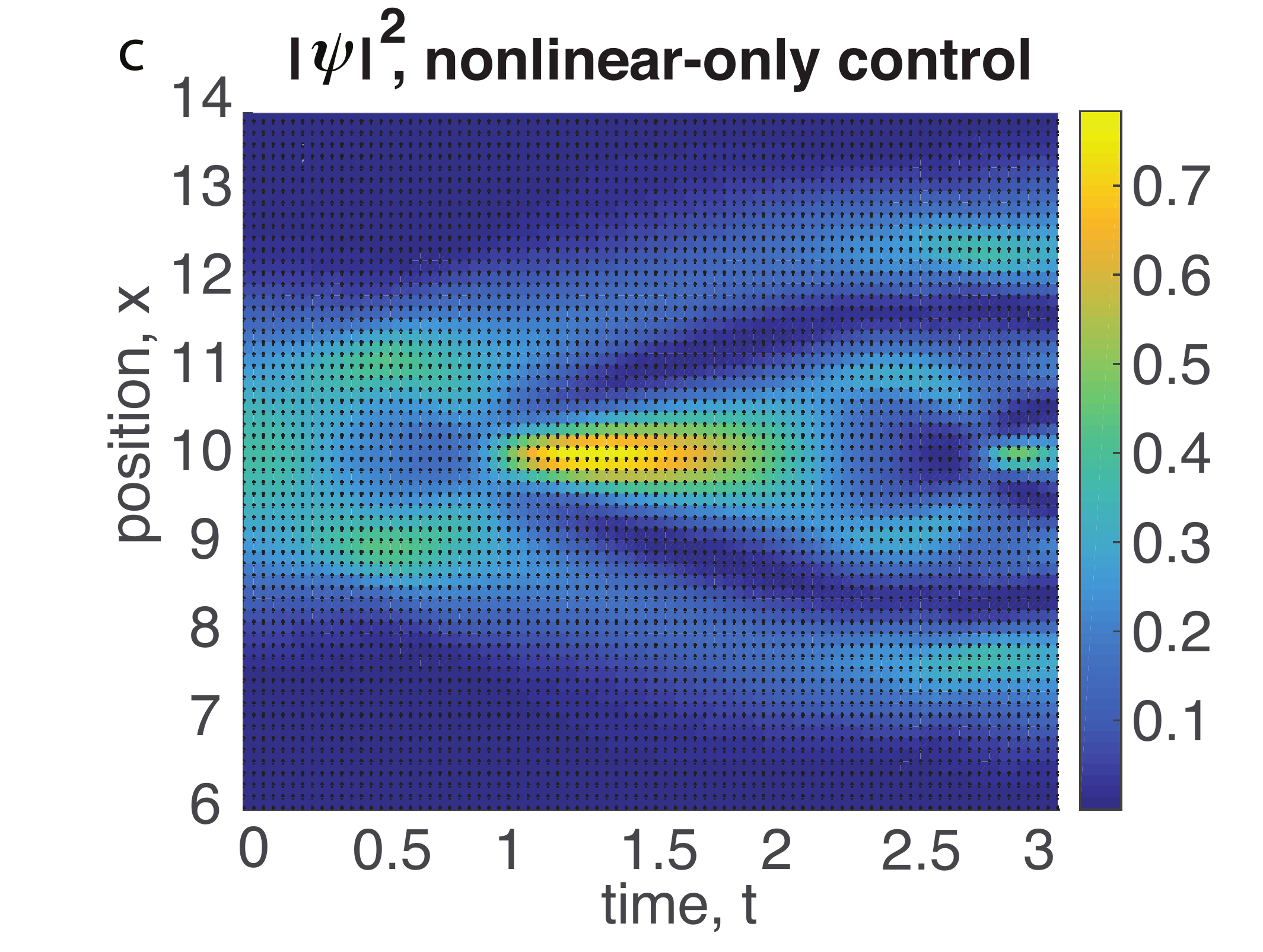}}
\subfloat{\includegraphics[width=0.26\textwidth, height=0.261\textheight, keepaspectratio]{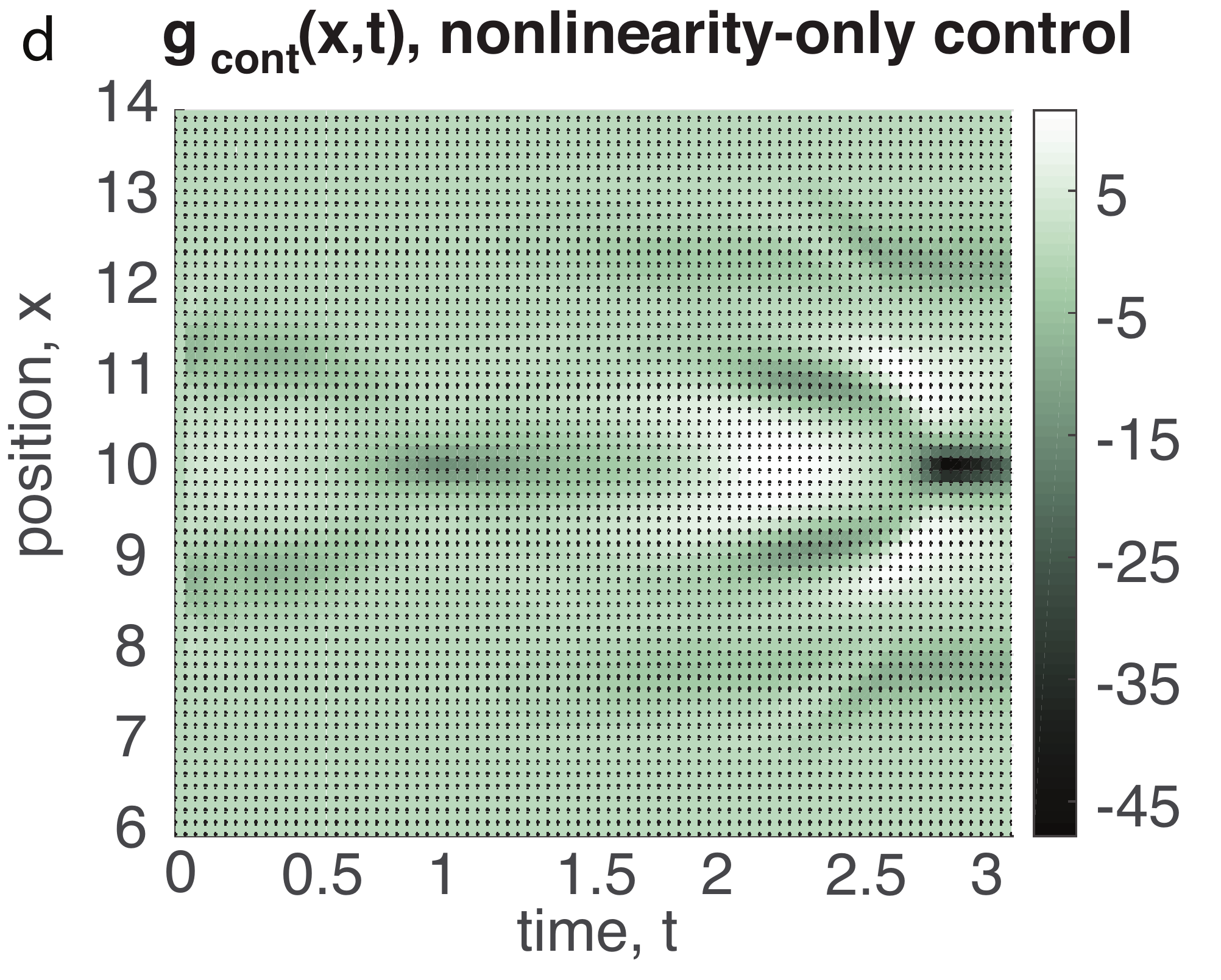}}
\end{center}
\begin{center}
\subfloat{\includegraphics[width=0.25\textwidth, height=0.25\textheight, keepaspectratio]{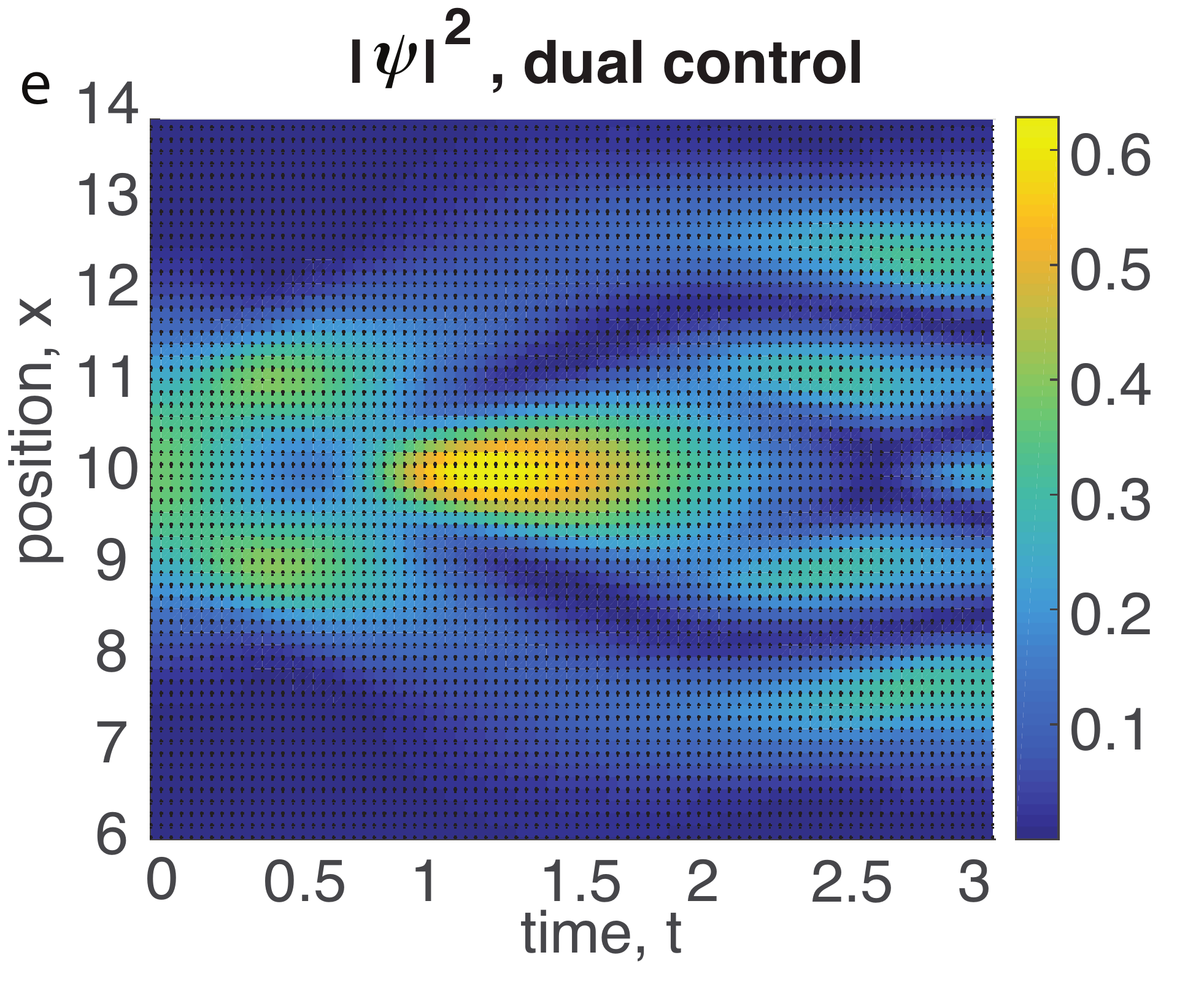}}
\subfloat{\includegraphics[width=0.25\textwidth, height=0.25\textheight, keepaspectratio]{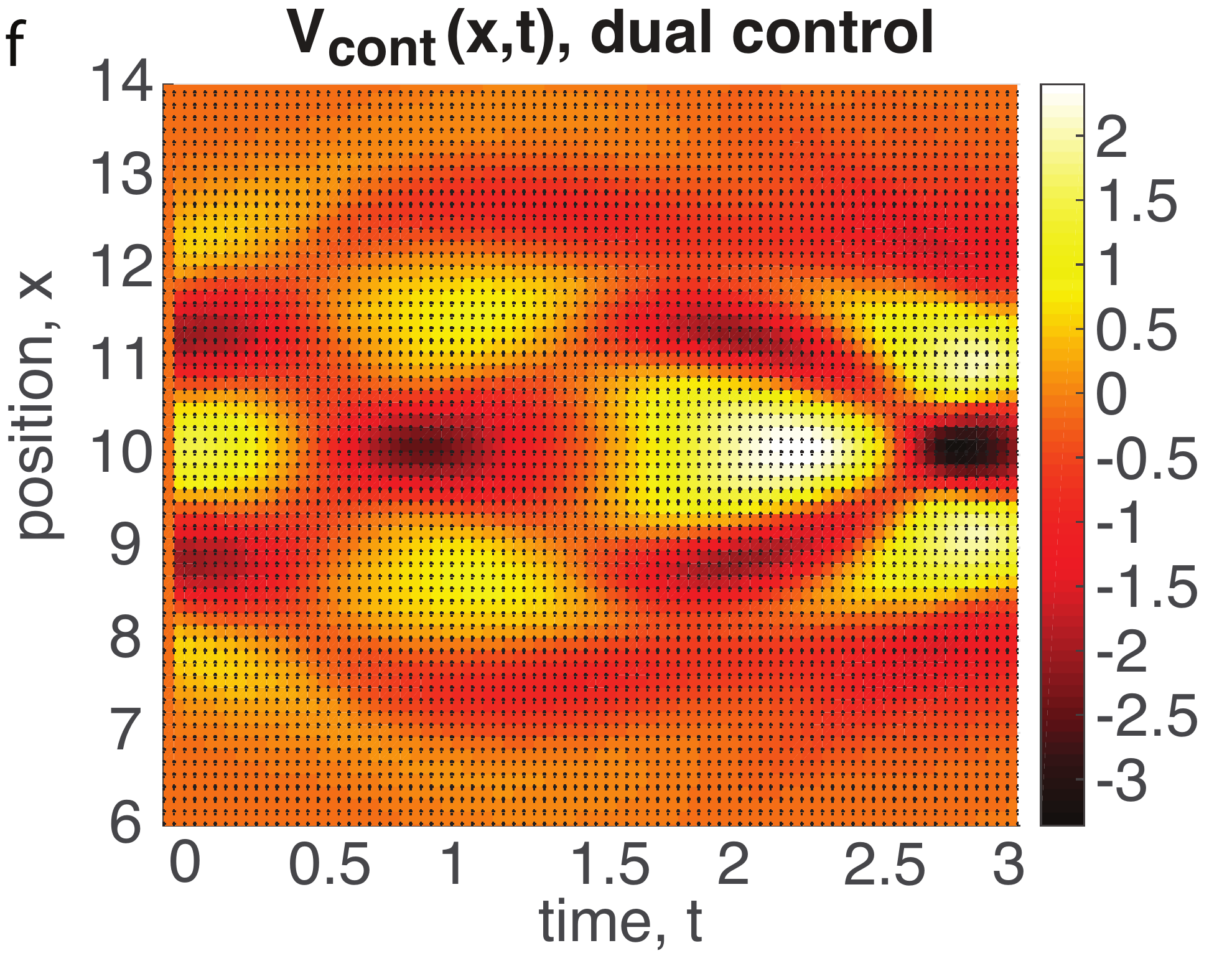}}
\subfloat{\includegraphics[width=0.25\textwidth, height=0.25\textheight, keepaspectratio]{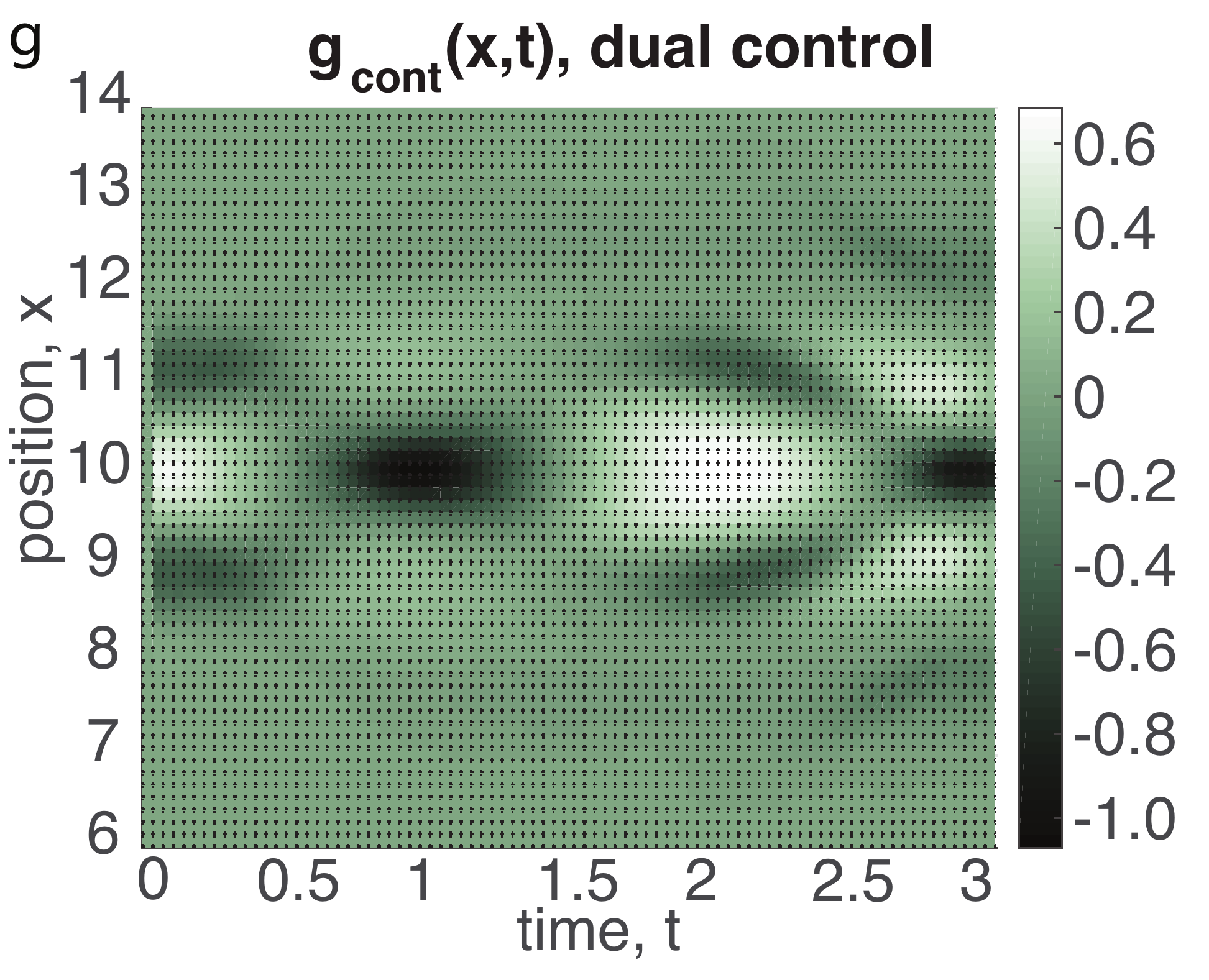}}
\end{center}
\caption{\label{fig:optim_results_psi} Optimal control results with $T=\pi$  for $P_{0 \rightarrow 4}$ for $g_0=5$. a),b) show the spatio-temporal dynamical evolution and control for potential-only optimization. c),d) similarly show results for nonlinearity-only control. e),f), and g) show the dynamics and both controls for dual control. Dual control mimics the dynamical behavior of potential-only control in a),b) by still heavily utilizing the potential as a primary control, while the nonlinearity-only scheme creates discernibly different dynamics.}
\end{figure*}

First, the short time results for $T=\pi$ are examined for each of the three control scenarios.  In this regime, the dynamics of the short propagation time result in strong-field control, and do not allow for resonant solutions in which the control has temporally and spatially oscillatory features at selected transition frequencies and wavenumbers related to the initial and final coherent modes.  As a representative example, the optimized mode evolution and controls (using initial guess $a=1/5\pi$, $\alpha(x)=\pi x/L$) for $g_0 = 5$ for the $P_{0 \rightarrow 4}$ transition are plotted in Figure \ref{fig:optim_results_psi}.  Potential-only control (Figures \ref{fig:optim_results_psi}a and  \ref{fig:optim_results_psi}b ) begins focusing the condensate early in the dynamics, and then splits the condensate. Utilizing only the nonlinearity as a control however, seems to take a different dynamical pathway with a large nonlinearity, as shown in Figures \ref{fig:optim_results_psi}c and \ref{fig:optim_results_psi}d, though also imparting an initial, but weaker focusing effect upon the condensate. Similar to potential-only control, the dual-control results of  Figures \ref{fig:optim_results_psi}e-g demonstrate that shaping of the nonlinearity primarily serves to assist $V_{\mathrm{cont}}$, which consequently makes the condensate dynamics similar to that of potential-only control in Figures \ref{fig:optim_results_psi}a and  \ref{fig:optim_results_psi}b.

For longer times, the resonant nature of controls become evident. In this situation we will follow standard practice of like analysis of the controlled dynamics with the linear Schr\"odinger equation. In particular we will examine selected illustrations of the spatial and temporal structure of the controls, the frequency and/or wavenumber features of the control upon Fourier Transform, as well as the mode population evolution over time. These simple tools applied to the cases of $V_{\mathrm{cont}}(x,t)$ and $g_{\mathrm{cont}}(x,t)$ give valuable insight into the control mechanisms. A more elaborate analysis by a Dyson-like expansion of the GPE is confounded by the nonlinear nature of the equation, and is beyond the scope of this work.

\begin{figure}[htb]
\begin{center}
\subfloat{\includegraphics[width=0.7\textwidth, height=0.7\textheight, keepaspectratio]{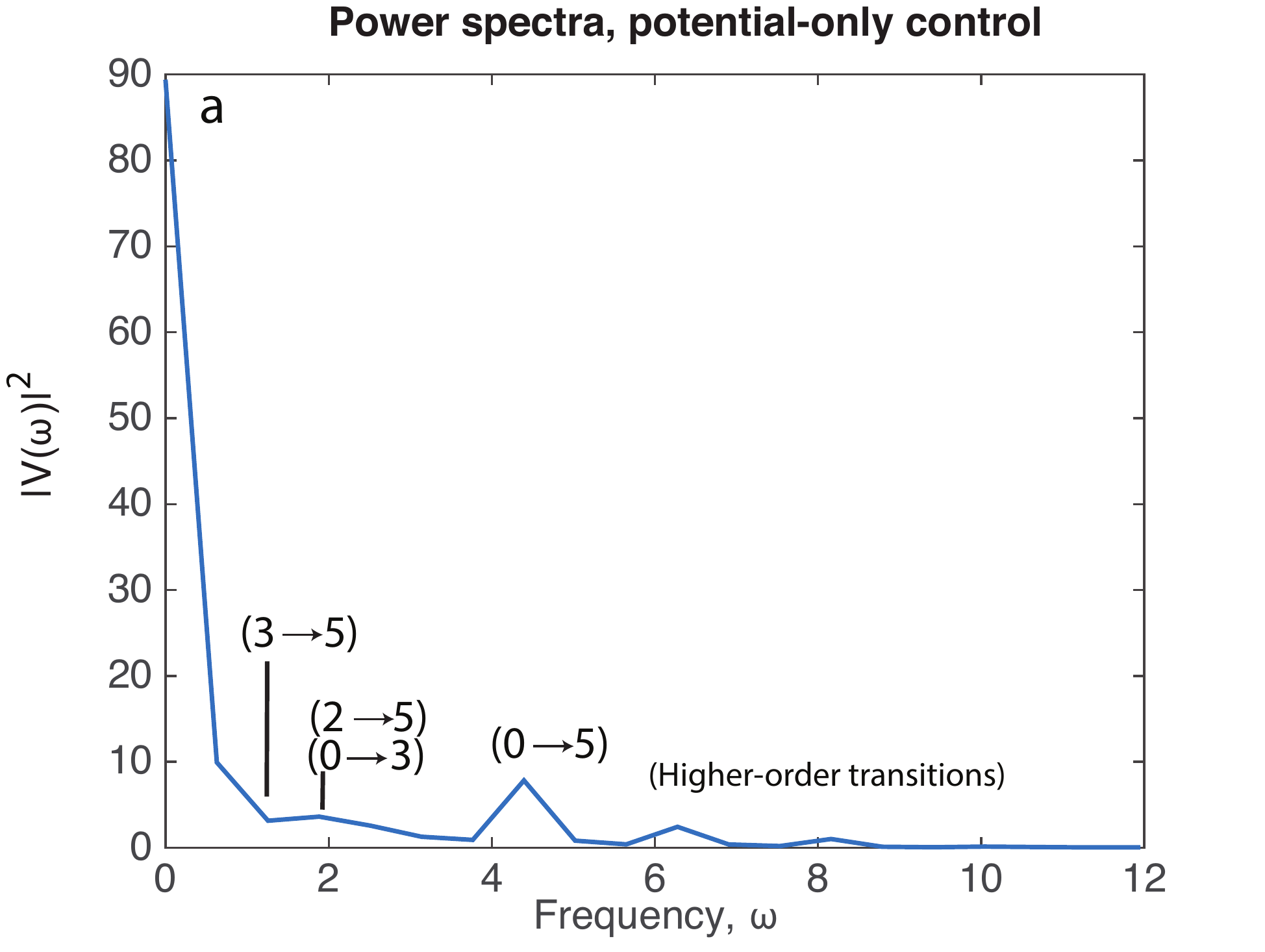}}

\subfloat{\includegraphics[width=0.7\textwidth, height=0.7\textheight, keepaspectratio]{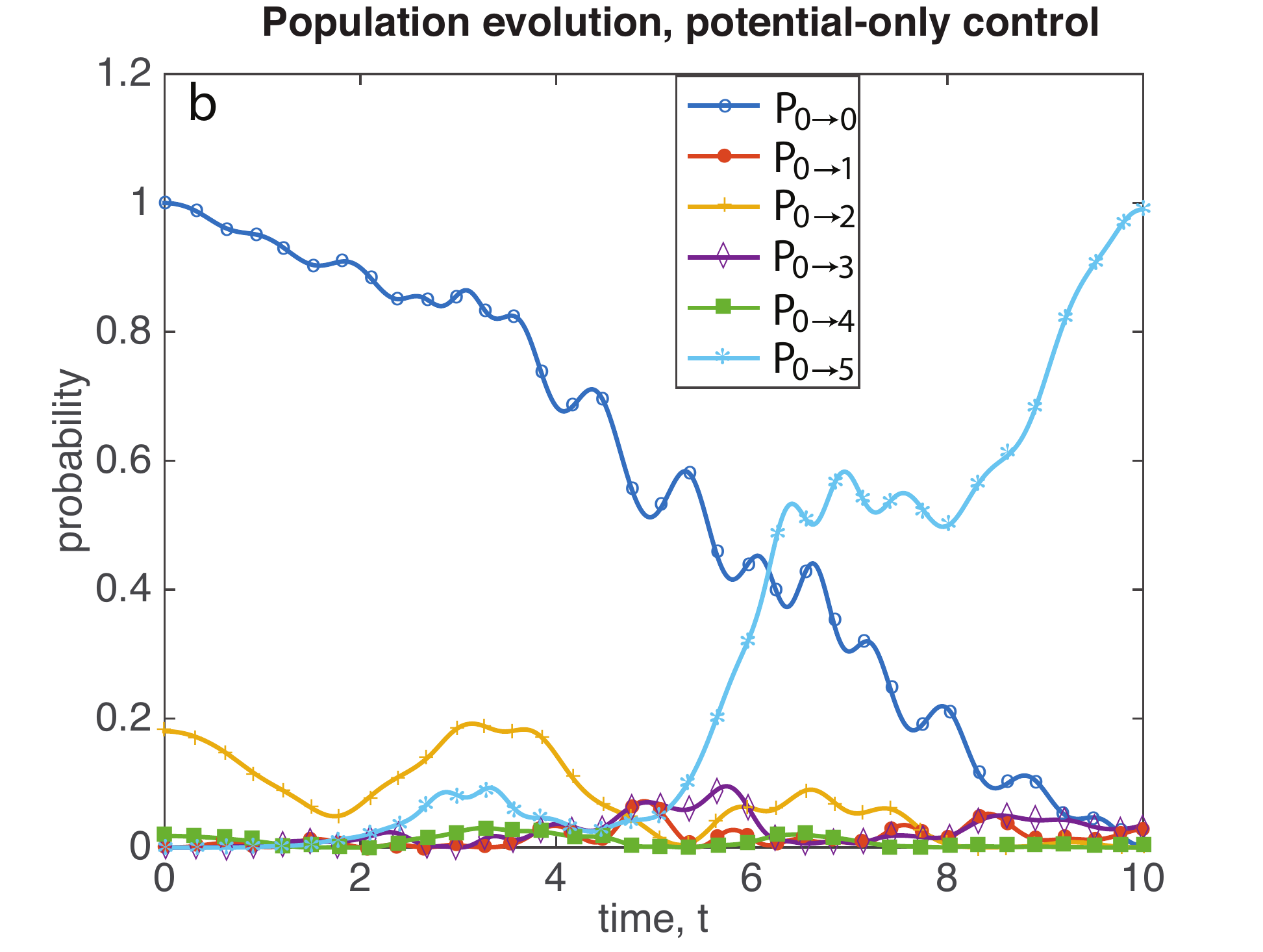}}
\caption{\label{fig:bec_g10_ts5_c1} Potential-only control with $g_0=10$. a) Spatially averaged potential power spectrum $|V(\omega)|^2$ and b) time-dependent population  $P_{0 \rightarrow j}(t)=|\langle \phi_j|\psi(t)\rangle |^2$  for the goal of the $P_{0 \rightarrow 5}$ transition at $T=10$. Frequencies $\omega > 5$ in a) are due to transitions into excited coherent modes higher than $\phi_5$. Given the coarse spectral resolution provided by simulated dynamics of duration $T=10$, the $2\rightarrow 5$ and $0 \rightarrow 3$ transitions are essentially degenerate in a).}
\end{center}
\end{figure}

Resonant control solutions will now be examined for $T=10$. We first consider $V_{\mathrm{cont}}(x,t)$, and to assess whether or not specific transitions between coherent modes are being directly excited, a spatially averaged power spectrum $|V(\omega)|^2$ arising from $V_{\text{cont}}(x,t)$ is computed from
\begin{align}
\label{eq:bec_powerspecpot}
V(\omega) &= \int_0^L \mathcal{F}_t[ V_{\text{cont}}(x,t)]  dx,
\end{align}
where $\mathcal{F}_t[f] = \int_0^T \exp(-i \omega t) f(t) dt$, and $|V(\omega)|^2$ shows the average spectral behavior of the control. The power spectrum is plotted in Figure \ref{fig:bec_g10_ts5_c1}a for  the $P_{0 \rightarrow 5}$ transition with potential-only control and $g_0=10$. The strong DC feature is a remnant of the initial controls in eq. (\ref{eq:V_init}). This leaves the remaining higher-frequency terms to be interpreted as relevant to the optimal dynamics of generating coherent mode transitions. Here, the $0\rightarrow 5$ transition at $\omega= 4.2$ is a prominent feature, but there also appear to be other transition energies excited from the potential, such as $3 \rightarrow 5$ ($\omega =1.8  $), which are likely involved in transitions involving intermediate modes. Higher frequency components are also present, suggesting that transitions into higher excited coherent modes also play a role in the transition pathway.  Some caution is called for in giving precise frequency interpretation to the spectral features, as power shifting can occur due to the controls. In this regard, the amplitude of the optimal controls often were not weak compared to the energy level spacing of the stationary modes (shown for $T=\pi$ cases in Fig. \ref{fig:optim_results_psi}, with similar results for cases where $T=10$). The time-dependent population dynamics between coherent modes shown in Figure \ref{fig:bec_g10_ts5_c1}b also indicate prominence of the direct $0 \rightarrow 5$ transition, as well as possible, indirect pathways $0 \rightarrow  3 \rightarrow 5$ and $0 \rightarrow 2 \rightarrow 5$.

\begin{figure}[htb]
\begin{center}
\subfloat{\includegraphics[width=0.68\textwidth, height=0.68\textheight, keepaspectratio]{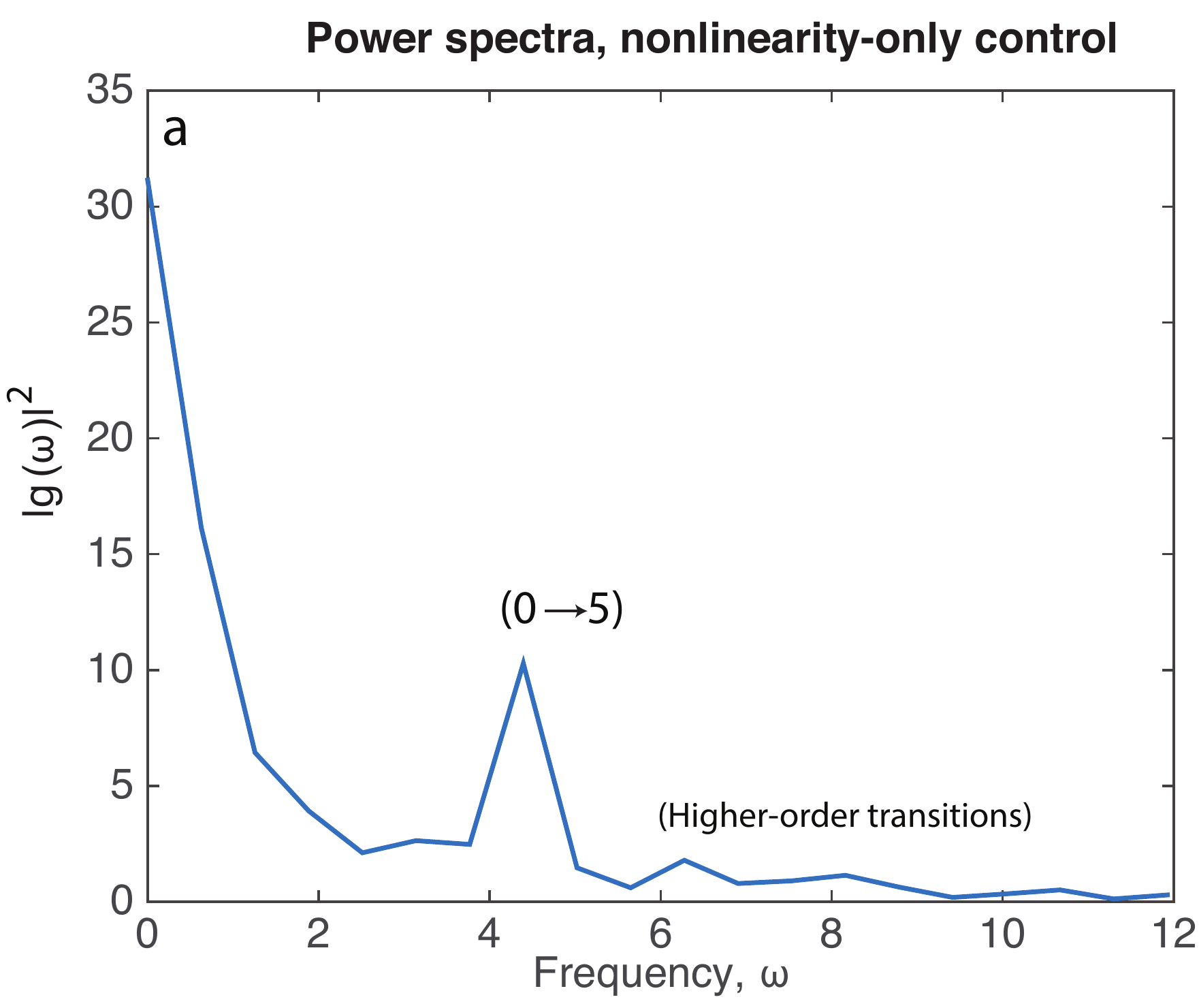}}

\subfloat{\includegraphics[width=0.75\textwidth, height=0.75\textheight, keepaspectratio]{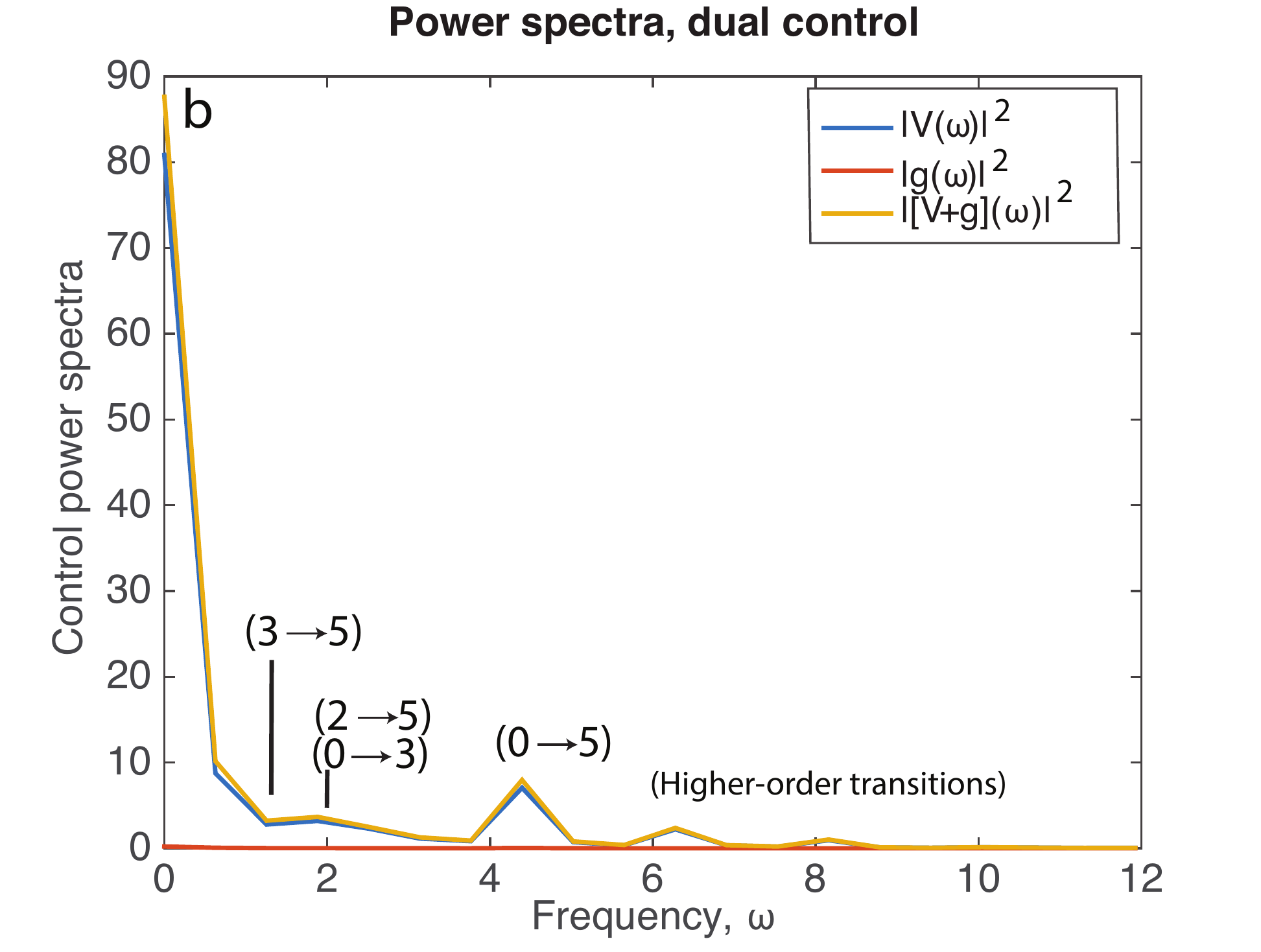}}
\caption{\label{fig:bec_g10_ts5_ch02} a) Spatially averaged nonlinearity-only control power spectrum $|g(\omega)|^2$ and b) dual control power spectra $|[V+g](\omega)|^2$ for the $P_{0 \rightarrow 5}$ transition, $g_0 = 10$. For dual control, $|V(\omega)|^2$ and $|g(\omega)|^2$ are also plotted to demonstrate that most of the energy from the controls resides in $V_{\mathrm{cont}}$. Again, frequencies $\omega > 5$ are likely due to higher-level excited mode transitions or power shifting.}
\end{center}
\end{figure}

Analogous to the potential power spectrum, a spatially averaged power spectrum $|g(\omega)|^2$ of the nonlinearity control can be calculated with
\begin{align}
\label{eq:bec_powerspecpot}
g(\omega) &= \int_0^L \mathcal{F}_t[g_{\mathrm{cont}}(x,t)|\psi(x,t)|^2]  dx,
\end{align}
as well as the spatially averaged power spectrum $|[V + g](\omega)|^2$ for dual potential and the nonlinearity control through
\begin{align}
\label{eq:bec_powerspecall}
[V + g](\omega) &= \int_0^L \mathcal{F}_t[ V_{\text{cont}} + g_{\mathrm{cont}}|\psi|^2 ]  dx.
\end{align}
The power spectrum $|g(\omega)|^2$ for nonlinear-only control is shown in Figure \ref{fig:bec_g10_ts5_ch02}a.  The $0 \rightarrow 5$ transition is again evident as the prominent feature. The results of dual control for the $P_{0 \rightarrow 5}$ transition are shown in Fig. \ref{fig:bec_g10_ts5_ch02}b. In addition to $|[V + g](\omega)|^2$, we also plot $|V(\omega)|^2$ and $|g(\omega)|^2$ in order to demonstrate that the bulk of the action on the system was provided by the potential, as $|[V + g](\omega)|^2$ and $|V(\omega)|^2$ are nearly identical. This is similar to the results for $T=\pi$ in Figure \ref{fig:optim_results_psi}, in which the dynamics of dual control closely resembled the results of potential-only control. Again, primarily direct transitions between initial and target modes occur.  High frequency components at $\omega \approx 6, 8$ are likely due to intermediate population of other excited coherent modes involving indirect transition pathways.

\begin{figure}[h]
\begin{center}
\subfloat{\includegraphics[width=\textwidth, height=\textheight, keepaspectratio]{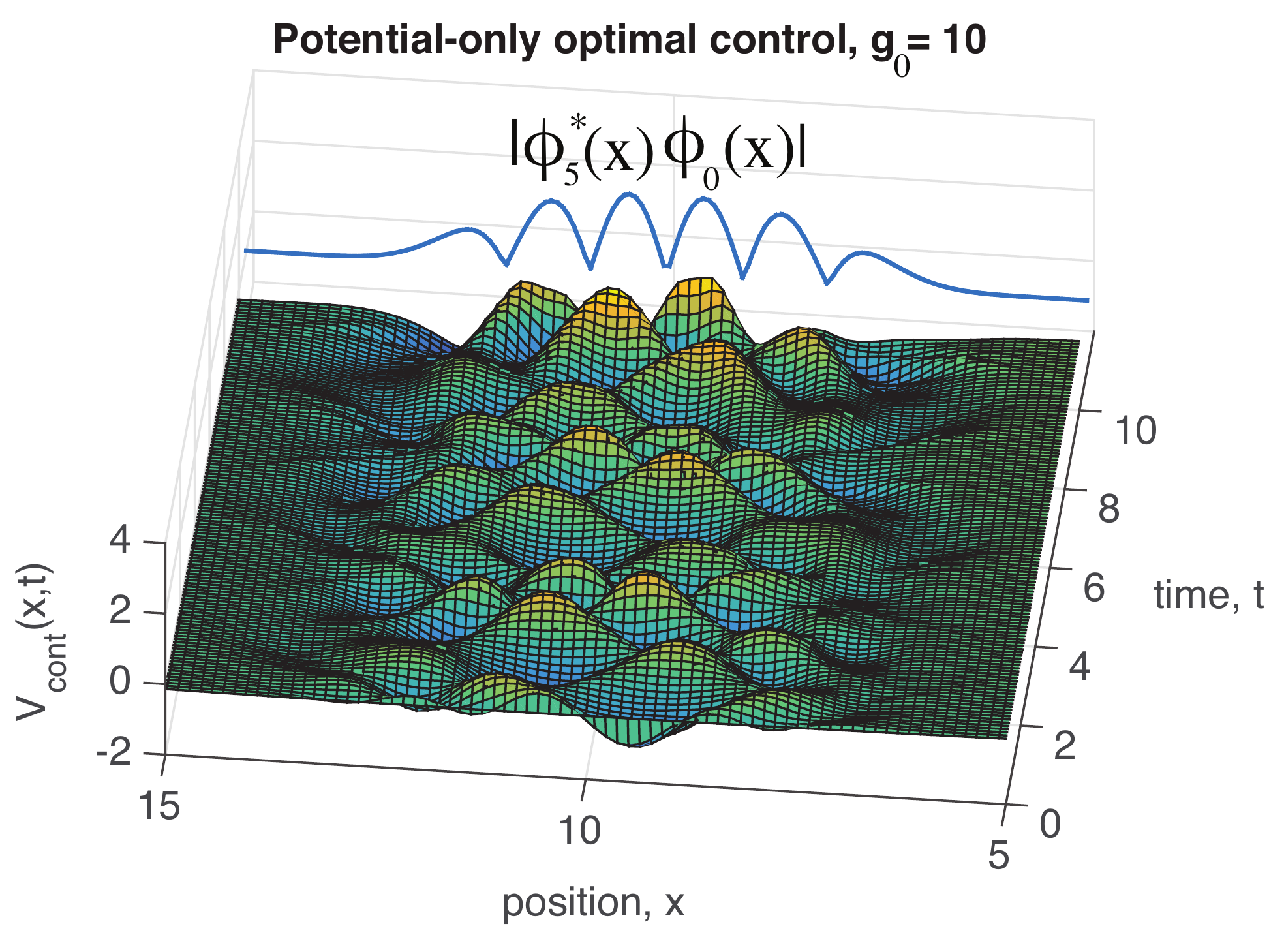}}
\caption{\label{fig:bec_spatial} Optimal control $V_{\mathrm{cont}}(x,t)$ for potential-only optimization of $P_{0\rightarrow 5}$ for $g_0=10$. The spatial profile of the potential follows the overlap of target and initial modes $|\phi_5^*(x)\phi_0(x)|$, but accompanied by complex temporal variations.}
\end{center}
\end{figure}

The goal of maximizing $|\langle \phi_f| \psi(T) \rangle |^2$ suggests that an effective control will reflect spatial features related to  related to $\sim \phi_f^*(x)\phi_0(x)$, which would be consistent with the dominant direct pathway found in Figs. \ref{fig:bec_g10_ts5_c1} and \ref{fig:bec_g10_ts5_ch02}. This spatial profile is plotted alongside the potential-only optimal controls for $g_0=10$ and $P_{0 \rightarrow 5}$ in Figure \ref{fig:bec_spatial} for comparison of the potential's evolving spatial structure. The spatial patterning of the control closely resembles the peaks occurring in $|\phi_5^*(x)\phi_0(x)|$, demonstrating that the optimal control adopts a similar form reminiscent of the latter pattern, but with the judicious variation of the spatial pattern over time to achieve the target transition. The other studied mode-to-mode transitions also showed similar behavior. 

\begin{figure}[htb]
\begin{center}
\subfloat{\includegraphics[width=\textwidth, height=\textheight, keepaspectratio]{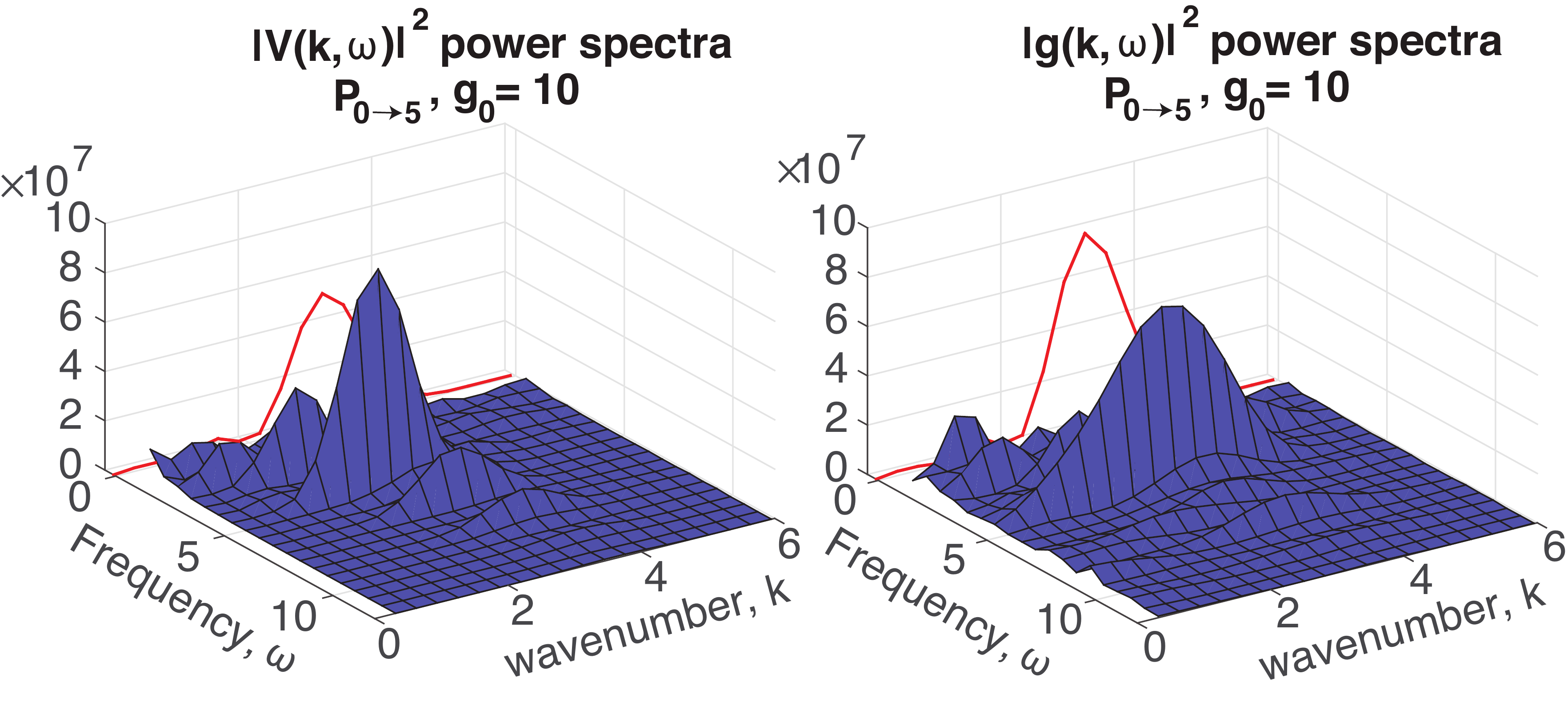}}
\caption{\label{fig:bec_2DFT} 2D power spectra of optimal controls for the $P_{0\rightarrow 5}$ transition and $g_0=10$. a) Power spectra for potential-only control. b) Power spectra for nonlinearity-only control. To aid visualization, DC components have been omitted. In both cases, the primary energy-level transition at $\omega = E_5-E_0 = 4.2$ is the leading feature of the spectra, although additional features also appear. The wavenumber contribution also exhibits similar shifts away from the expected structure (\emph{i.e.} $|\mathcal{F}_x[\phi_5^*(x)\phi_0(x)]|^2$, scaled and displayed as a solid line). These variations reflect the likely complex interplay of the spatio-temporal roles of the control, particularly when it is a strongly contributing term in the Hamiltonian.   }
\end{center}
\end{figure}

Examining the joint spatial and temporal features provides guiding insight into optimal controls, and a complementary picture is the 2D spectra for controls in the frequency $\omega$ and wavenumber $k$ domains, given respectively for potential-only and nonlinear-only controls by
\begin{align}
\label{eq:2dft}
V(k,\omega) &= \mathcal{F}_{x,t}[V_{\mathrm{cont}}(x,t)] \\
g(k,\omega) &= \mathcal{F}_{x,t}[g_{\mathrm{cont}}(x,t) |\psi(x,t)|^2 ] \\
\label{eq:bec_2dft}
\mathcal{F}_{x,t}[f] &=  \int_0^T \int_0^Le^{- i \omega t} e^{-2\pi i k x} f(x,t) dx dt.
\end{align}
These spectra $|V(k,\omega)|^2$ and $|g(k,\omega)|^2$ are plotted for the $P_{0 \rightarrow 5}$  transition with $g_0 = 10$ and $T=10$ in Figure  \ref{fig:bec_2DFT}. In both simulations, the primary energy level transition located at $\omega = E_5-E_0$ is the prominent feature, although it can be seen that higher energy modes in these particular control solutions are linked to higher wavenumber components. In an analogous fashion to the expected spatial structure displayed in Figure \ref{fig:bec_spatial}, the spectra maintains some overlap with the expected wavenumber structure given by $|\mathcal{F}_x[\phi_5^*(x)\phi_0(x)]|^2$, denoted by the curve in Figure  \ref{fig:bec_2DFT} . However, there is a clear deviation present from the profile given by $|\mathcal{F}_x[\phi_5^*(x)\phi_0(x)]|^2$. This deviation was observed most strongly for higher $g_0$ values for each target mode transition, which highlights the limited view of a mechanism analysis based on just examining the controls; the finding is parallel with like analyses of controlled dynamics with the linear Schr\"odinger equation. Again, for any given targeted mode transition there is expected to be many optimal controls, with a diversity of temporal and spatial structural features. The results shown here display a sampling of possible motifs that may be encountered under the rich GPE dynamics.

 \section{Conclusion}
 \label{sec:becconclusion}
BECs possess a unique opportunity to explore and exploit the nonlinear features driving their dynamics, and here we demonstrated that this control can be successfully used to drive transitions between nonlinear stationary modes. These nonlinear coherent modes are analogous to eigenstates of a linear Schr\"odinger equation, and arbitrary order excited modes were successfully characterized through an imaginary time propagation method that adaptively altered the spectrum of the Hamiltonian to locate specific excited modes. This S-AITP method was successful at locating coherent modes in a harmonic trap, and exploring its general capability for asymmetric trapping potentials will further establish its utility as a characterization method of nonlinear Hamiltonian eigenmodes. 

The perspective taken in this work rests on unleashing the spatial-temporal nature of the controls in order to assess performance with the widest available freedom. Although the freedom in mode-to-mode optimization should permit many solutions to exist, the particular solutions examined here revealed a leading, direct transition between initial and target modes. When operating with dual controls using both the potential and the nonlinear term, the solutions were dominated by the potential. Finally, we hope that this work opens further study of the rich domain of spatio-temporal control possibilities for manipulating BEC dynamics, which may enable access to new fundamental and practical domains of its experimental implementation.

\begin{acknowledgements}
The authors would like to thank Benjamin Russell, Tak-San Ho, Ruixing Long, and Ulrich Hohenester for their insightful conversations. D.H. acknowledges support from NSF fellowship (DGE 1148900), and H.R. acknowledges support from NSF (CHE-1464569) for landscape principles and ARO-MURI (W911NF-11-1-0268) for quantum information science.
\end{acknowledgements}

\bibliographystyle{apsrev4-1}
\bibliography{hocker}	

\end{document}